# Attoheat transport phenomena

J. Marciak-Kozlowska[1], M. Pelc[2], M. A. Kozlowski[3]

[1]Institute of Electron Technology, Warsaw Poland

[2]Institute of Physics , Maria Sklodowska-Curie University, Lublin, Poland

[3] Kabaty, Warsaw, Poland

## Abstract

Fascinating developments in optical pulse engineering over the last 20 years lead to the generation of laser pulses as short as few femtosecond, providing a unique tool for high resolution time domain spectroscopy. However , a number of the processes in nature evolve with characteristic times of the order of 1 fs or even shorter. Time domain studies of such processes require at first place sub-fs resolution, offered by pulse depicting attosecond localization. The generation, characterization and proof of principle applications of such pulses is the target of the attoscience.

In the paper the thermal processes on the attosecond scale are described. The Klein-Gordon and Proca equations are developed. The relativistic effects in the heat transport on nanoscale are discussed. It is shown that the standard Fourier equation can not be valid for the transport phenomena induced by attosecond laser pulses. The heat transport in nanoparticles and nanotubules is investigated

CONTENTS

# Chapter 1

# Overview of the research

1.1 HEDS

In the present state of development the quantum mechanics (QM) is not complete enough to treat all the precise details of the motion of individual elementary particles: electrons and photons.

To treat those details (e.g. entanglement) we should have to go to some as yet unknown deeper level, which has the same relationship to the atomic level as the atomic level has to that of Brownian motion.

To overcome the QM limitations we should to take advantage of properties of matter that depended significantly on the subquantum level. One way to do this would be to make our observations with the aid of processes that were very fast compared with the subquantum fluctuation (Brownian motion), so that the whole measurement would be over before these fluctuations could have changed things very much. Such rapid processes are most likely to be obtained with the attosecond laser pulses.

The seminal paper: *Single – Cycle Nonlinear Optics* (*Science*, vol. **320** (2008) 1614) published by *prof. Ferenc Krausz* at Max Planck Institute für Quantenoptik Garchhing , for the first time presents the method for the production of the laser pulses shorter than 100 as. Considering that the "electron year" in hydrogen atom is of the order of 100 as ($T = \frac{2\pi r_B}{\alpha c}$, $r_B$ = Bohr radius, $a$ = fine structure constant, $c$ = vacuum light speed) we can observe the trajectory (*a la* Bohm) of the electron moved in the field of the *virtual electron – positon pairs* (*life time* of virtual pairs is of the order of 100 as).

With the attosecond laser pulses we can study the the relaxation phenomena on the subatomic quantum level. The *quantum relaxation* time is equal $\tau = h/m a^2 c^2$, where $h$ is the Planck constant, $m$ is the electron mass, and $a = 1/137$ is the fine structure constant. For electrons the $\tau$ is of the order 100 as. As it was shown in monograph: *Thermal processes using attosecond laser pulses*: M.Kozlowski, J Marciak-Kozlowska, Spinger 2006, for pulses shorter than $\tau$ the master equation for thermal processes changes the structure from *parabolic* to *hyperbolic partial differential equations*. . Considering that the *Schrodinger equation* is the *parabolic equation*, for $\Delta t < \tau$ we argue that for attosecond laser pulses hyperbolicity of transport processes leads to the new more *general quantum hyperbolic* equation ( instead of the Schrodinger equation)

The increase in light intensity available in the laboratory over the previous 20 years has been astounding. Laser peak power has climbed from giga-watts to petawatts in this time span, and accessible focused intensity has increased by at least seven orders of magnitude. Such a dramatic increase in light brightness has accessed an entirely new set of phenomena. High repetition rate table top lasers can routinely produce intensity in excess of $10^{19} W/cm^2$, and intensities of up to $10^{20} W/cm^2$ are possible with the latest petawatt class systems (Mourou et al., 1998; Mourou and Umstadter, 2002; Perry and Mourou, 1994). Light-matter interactions with single atoms are strongly non-perturbative and electron energies are relativistic. The intrinsic energy density of these focused pulses is very high, exceeding a gigajoule per $cm^3$. The interactions of such intense light with matter lead to dramatic effects, such as high temperature plasma creation, bright X-ray pulse generation, fusion plasma production, relativistic particle acceleration, and highly charged ion production (Mourou and Umstadter, 2002).

Such exotic laser-matter interactions have led to an interesting set of applications in high field science, and high energy density physics (HED physics). These applications span basic science and extend into unexpected new areas such as fusion energy development and astrophysics. In this paper some of these new applications will be reviewed. The topics covered here do not represent a comprehensive list of applications made possible with high intensity short pulse lasers, but they do give a flavor of the diverse areas affected by the latest laser technology. Most of the applications discussed here are based on recent experiments using lasers with peak power of 5-100 TW. Many important leaps in laser technology have driven the rapid advances in HED and high field science over the last 15 years. The enabling advancement for this technological progress was the invention of chirped pulse amplification (CPA) lasers (Strickland and Mourou, 1985). A broad bandwidth, mode-locked laser produces a low power, ultrafast pulse of light, usually with duration of 20-500 fs. This short pulse is first stretched in time by a factor of around ten thousand from its original duration using diffraction gratings. This allows the pulses, now of much lower peak power, to be safely amplified in the laser, avoiding the deleterious nonlinear effects which would occur if the pulses had higher peak power (Koechner, 1996). These amplified pulses are, finally, recompressed in time (again using gratings), in a manner that preserves the phase relationship between the component frequencies in the pulse. The CPA output has a duration near that of the original pulse but with an energy greater by the amplification factor. In high-energy CPA systems, severe nonlinearities occurring when the pulse propagates in air can be a major problem, so the pulse must be recompressed in an evacuated chamber.

The first generation of CPA lasers were based mainly on flashlamp pumped Nd:glass amplifiers (Danson et al., 1998; KJtagawa et al., 2002; Patterson and Perry, 1991; Strickland and Mourou, 1985). These glass based lasers are usually limited to pulse duration of greater than about 300 fs because of gain narrowing in the amplifiers (Blanchot et al., 1995). The most significant scaling of this approach to CPA was demonstrated by the Petawatt laser at Lawrence Livermore National Laboratory in the late 1990s (Perry et al., 1999). This laser demonstrated the production of 500 J per pulse energy with duration of under 500 fs yielding over $10^{15}$W of peak power. Since this demonstration, a number of petawatt laser projects have been undertaken around the world (Service, 2003).

The second common approach to CPA uses Ti:saphire as the amplifier material. This material permits amplification of much shorter pulse durations often down to 30 fs.

However, the short excited state lifetime of Ti:sapphire (~3us) requires that the material be pumped by a second laser (usually a frequency doubled Nd:YAG or Nd:glass laser). The inherent inefficiencies of this two step pumping usually limit the output energy of such a laser to under a few joules of energy per pulse. A number of multi-terawatt lasers based on T:sapphire now operate in many high intensity laser labs world wide (Yamakawa et al., 1999; Walker et al., 1999; Patterson et al., 1999). An example of a typical table top scale multi-terawatt system is the laser we have constructed at the University of Texas. This optical schematic of this system, which delivers 0.75 J in 35 fs pulses at a 10 Hz repetition rate, is illustrated in Fig. 2. To date, the largest scaling of T:sapphire technology has been to the 100 TW power level (Yamakawa and Barty, 2000; Patterson et al., 1999) with energy of up to 10 J per pulse. Scaling of Tirsapphire to the petawatt power level is being pursued in Japan and will likely be achieved within a year or so (Service, 2003).

The third major technology upon which a new generation of high peak power CPA lasers is being based is optical parametric chirped pulse amplification (OPCPA) (Ross et al., 1997). In this approach, amplification of the stretched pulses occurs not with an energy storage medium like Ndrglass or Tirsapphire but via parametric interactions in a nonlinear crystal. This approach is quite attractive because of the very high gain per stage possible (often in excess of $10^4$ per pass) and the very broad gain bandwidth possible in principal. To date, a number of CPA demonstrations with OPCPA have been published (Ross et al., 2000; Jovanovic et al., 2002; Leng et al., 2003). Though there are still significant technological issues to be resolved with OPCPA, this technique promises to lead to a new generation of high peak power, femtosecond lasers.

The physics accessible with this class of lasers is quite extreme. The science applications made possible with these extremes can be simply classified into two categories. First, by temporally compressing the pulses and focusing to spots of a few wavelengths in diameter, these lasers concentrate energy in a very small volume. A multi-TW laser focused to a few microns has an intrinsic energy density of over $10^9 J/cm^3$. This corresponds to about l0keV of energy per atom in a material at solid density. As a result, quite high temperatures can be obtained. Such energy density corresponds to pressure in excess of l0Gbar. Many applications of ultraintense lasers stem from the ability to concentrate energy to high energy-density which can lead to quite extreme states of matter.

The second class of applications arises from the high field strengths associated with a very intense laser pulse. At an intensity of over $10^{18}$W/cm$^2$, an intensity quite easily achievable with modern table top terawatt lasers, the electric field of the laser exceeds ten atomic units, over $10^{10}$V/cm. Consequently, the strong field rapidly ionizes the atoms and molecules during a few laser cycles. At intensity approaching $10^{21}$ W/cm$^2$, the current state of the art with petawatt class lasers, the electric field is comparable to the field felt by a K-shell electron in mid Z elements such as argon. The magnetic field is nearly 1000 T. An electron quivering in such a strong electric field will be accelerated to many MeV of energy in a single optical cycle. Such charged particles will experience a very strong forward directed force resulting from the Lorentz force.

High peak power femtosecond lasers are unique in their ability to concentrate energy in a small volume. A dramatic consequence of this concentration of energy is the ability to create matter at high temperature and pressure. Matter with temperature and density near the center of dense stars can be created in the laboratory with the latest high intensity lasers. For example, solid density matter can be heated to temperature of over keV. Under these conditions, the particle pressure inside the sample is over 1 billion atmospheres, far higher than any other pressure found naturally on or in the earth and approaches pressures created in nuclear weapons and inertial confinement fusion implosions.

Study of the properties of matter at these extreme conditions, namely higher in the temperature range of 1-1000 eV, is crucial to understanding many diverse phenomenon, such as the structure of planetary and stellar interiors or how controlled nuclear fusion implosions (inertial confinement fusion or ICF) evolve. Yet, despite the wide technological and astrophysical applications, a true, complete understanding of matter in this regime is not in hand. A large obstacle is posed by the fact that theoretical models of this kind of matter are difficult to formulate. While the atoms in these warm and hot dense plasmas are strongly ionized, the very strong coupling of the plasma, and continuum lowering in the plasma dramatically complicates traditional plasma models which depend on two body collision kinetics (Spitzer, 1967). Even the question of whether electrons in this state are free or bound is not as clear cut as it is in a diffuse plasma.

Over the last 15 years, since the development of high energy short pulse lasers, there have been a number of experimental studies aimed at isochorically heating solid density targets with a femtosecond laser pulse (Audebert et al., 2002; Widmann et al., 2001). This class of experiments uses a femtosecond pulse to heat, at solid density, an inertially confined target on a time scale much faster than the hot material can expand by hydrodynamic pressure. This approach can be very powerful. Not only can spectroscopic diagnostics be implemented to derive information on the heated material (such as ionization state) (Nantel et al., 1998) but isochoric heating experiments can also enable laser heated pump-probe experiments. Pump-probe experiments can use an optical, or laser-generated X-ray pulse to probe the material on a time scale before it can expand. This yields, for example, conductivity information through reflectivity and transmission probing (Widmann et al., 2001), as well as XUV and X-ray opacity if a short wavelength probe is employed (Workman et al., 1997). The laser heats a thin slab of material, with thickness usually comparable to or less than an optical skin depth. When the material is heated from its initially cold, solid state, the plasma will be in the "strongly coupled" regime1, even at temperature approaching 1 keV.
. Direct laser heating has drawbacks. Because of the small depth of such a skin depth, the expansion time of the sample is often only 100 or 200 fs. The few hundred fs release time of the heated material is comparable to the electron-ion equilibration time (Spitzer, 1967) and hampers interpretation of this kind of experiment. Microsopically rough surface finish of the very thin target often leads to a large uncertainty in the initial density, complicating analysis of the data.

## 1.2 Wakefield accelerators

The plasma particle accelerators has received a considerable interest in the past decade. Improvements on the laser technology and laboratory facilities made the plasma particle accelerators a strong alternative for the huge high-energy colliders. One sort of the plasma particle accelerators is called as the laser wakefield accelerators (LWFA), in which a short, intense laser pulse is used to create a wake in the plasma. Current systems of high power laser pulses produce a plasma wave which can accelerate elections to the energies of one trillion electron volts in a distance less man a centimeter.

When a short, intense laser pulse is sent to plasma, it generates a plasma wave whose amplitude is larger man mat of the wave produced by the single laser pulse using the same total energy. Another series of pulses is used to push extra electrons into the wave, which accelerates them into a narrowly focused beam at nearly the speed of light However, the problem is basically interaction of electromagnetic wave with particles, properties of the medium in which the particles present has great importance. The limits on die LWFA due to properties of the plasma are discussed.

The plasma fluctuations while the laser pulse propagates inside are described by the fluid equations. Assuming that the plasma is cold, and there is no background magnetic field, the motion of electrons, taking the background ions are stationary, is described by

$$\frac{\partial p}{\partial t} + V \cdot \nabla_p + \nabla P = -e\left[E + \left(\frac{V}{c} \times B\right)\right] \tag{1.9}$$

where v and $n$ are die velocity and the density of die electrons, respectively, $P$ is the pressure. Furthermore, the equation of continuity and the Poisson's equation are employed.

$$\frac{\partial n}{\partial t} + \nabla \cdot (nV) = 0 \tag{1.10}$$

$$\nabla \vec{E} = 4\pi\rho \tag{1.11}$$

After defining the momentum *p*, as the product of mass, velocity, and the number density, linearization of the velocity and density together with taking the ponderomotive force into consideration yields

$$m\frac{\partial V}{\partial t} + m v_{ei} V = -e(\nabla \Phi + \nabla \Phi_{NL}) \tag{1.12}$$

substituting the Poisson's equation into Eq. (1.12), a second-order partial differential equation for the density is obtained:

$$\frac{\partial^2 n}{\partial t^2} + v_{ei}\frac{\partial n}{\partial t} + n\omega_p^2 = -\frac{\omega_p^2}{2\pi e}\nabla^2 \Phi_{NL} \tag{1.13}$$

where $\omega_p$ is the plasma frequency, $v_{ei}$ is the electron ion collision frequency, and $\Phi_{NL}$ is the nonlinear potential for the ponderomotive force of the laser beam. The Poisson's equation is used once more to define the number density in terms of the electrostatic potential. Having introduced this fact, the fluctuations in the plasma will be totally described by an inhomogeneous differential equation for the electrostatic potential. The nonlinear potential is well defined in terms of the normalized vector potential *a* such as $\Phi_{NL}$ = -(mc²/2e)|a²(r,z,t) Assuming the laser beam profile as bi-Gaussian and inducing the degrees of freedom of the problem by combining the time dependency and the longitudinal displacement together such that defining a time-dependent displacement ξ= *z-v$_p$t*, where *v$_p$* is the phase velocity of the exerted plasma wave the nonlinear potential is then described as

$$\frac{\partial^2 \Phi}{\partial t^2} + v_{ei}\frac{\partial \Phi}{\partial t} + \Phi\omega_p^2 = -\omega_p^2 \Phi_{NL} \tag{1.14}$$

Here the parameters $a_z$ and $a_r$ are rms pulse length and spot size, respectively. Thus, Eq. (1.14) turns out to be

$$\Phi_{NL} = -\frac{mc^2}{2e} a_0^2 e^{-[(2r^2/\sigma_r^2)-(\xi^2/\sigma_z^2)]} \tag{1.15}$$

where a=vei/2v$_p$, and k$_p$=ωp/v$_p$. Solving Eq. (1.15) with $k^2=kp^2-a^2$, yields

$$\frac{\partial^2 \Phi(r,\xi)}{\partial y^2} + 2\alpha \frac{\partial \Phi(r,\xi)}{\partial \xi} + k_p^2 \Phi(r,\xi) = -\frac{mk_p^2 c^2}{2e} a_0^2 e^{-[(2r^2/\sigma_r^2)+(\xi^2/\sigma_z^2)]} \qquad (1.16)$$

The partial derivatives of Eq. (1.16) describes the so-called wakefields along the longitudinal and the radial directions. Making use of the error functions and the limits $\xi \to -\infty$ these wakefields are expressed as

$$E_z(r,\xi) = \frac{\sqrt{\pi}}{4e} mc^2 \sigma_z k_p^2 e^{-[(k^2\sigma_r^2/4)-(r^2\sigma_r^2)]} \cdot \cos k\xi \qquad (1.17)$$

$$E_r(r,\xi) = -\sqrt{\pi} \frac{mc^2}{e} a_0^2 \frac{r}{\sigma_r^2} \sigma_z e^{-(2r^2/\sigma_r^2)+(k_p^2\sigma_z^2)} \cdot \sin k\xi \qquad (1.18)$$

The axial wakefield reaches the maximum amplitude at $\xi$ =(n Pi/k),(n =0,1,2,...). Naturally, these extremes alternate depending on the integer n, such that Ez has a maximum for even n and it has a minimum for odd n on the z axis. More clearly it has a maximum at z= $v_p$t, (2Pi/k + $v_p$t,.... and it has a minimum at points z=Pi/k+$v_p$t,...

1.3  High Energy Heat Transport and Special Relativity Theory

In the interaction of laser beam with thin solid target, fast ion beams and electrons are generated with large density. When the energy of laser beam is highly enough the ion and electron velocities are relativistic and transport of thermal energy must be described by relativistic equations.

In the subsequent we develop the description of the heat transport in Minkowski spacetime. In the context special relativity theory we investigate the Fourier equation and hyperbolic diffusion equation. We calculate the speeds of the heat diffusion in Fourier approximation and show that for high energy laser beam the heat diffusion exceeds the light velocity. We show that this results breaks the causality of the thermal phenomena in Minkowski spacetime. The same phenomena we describe in  the framework of hyperbolic heat diffusion equation and show that in that case speed of diffusion is always smaller than light velocity.

We may use the concept that the speed of light *in vacuo* provides an upper limit on the speed with which a signal can travel between two events to establish whether or not any two events could be connected. In the interest of simplicity we shall work with one space dimension $x_1 = x$ and the time dimension $x_o = ct$ of the Minkowski spacetime. Let us consider events (1) and (2): their Minkowski interval $\Delta s$ satisfies the relationship:

$$\Delta s^2 = c^2 \Delta t^2 - \Delta x^2 \qquad (1.19)$$

Without loss of generality we take Event 1 to be at $x = 0$, $t = 0$. Then Event 2 can be only related to Event 1 if it is possible for a signal traveling at the speed of light, to connect them. We illustrate three general possibilities in Fig. 1, where Event 2 is at ($\Delta x$, $c\Delta t$), its relationship to Event 1 depending on whether $\Delta s > 0, = 0,$ or $< 0$.

We may summarize the three possibilities as follows:

Case A  *timelike interval*, $|\Delta x_A| < c\Delta t$, or $\Delta s^2 > 0$. Event 2 can be related to Event 1, events 1 and 2 can be in causal relation.

Case B  *lightlike interval*, $|\Delta x_B| = c\Delta t$, or $\Delta s^2 = 0$. Event 2 can only be related to Event 1 by a light signal.

Case C  *spacelike interval* $|\Delta x_A| > c\Delta t$, or $\Delta s^2 < 0$. Event 2 cannot be related to Event 1, for in that case $v > c$.

Now let us consider the case C in more details. At first sight it seems that in case C we can find out the reference frame in which two Events $c^>$ and $c^<$ always fulfils the relations $t_{c^>} - t_{c^<} > 0$. but it is not true. For lest us choose the inertial frame U' in which $t_{c^>} - t_{c^<} > 0$. In reference frame U which is moving with speed $V$ relative to U', where

$$V = c \frac{c(t'_{c^>} - t'_{c^<})}{x'_{c^<} - x'_{c^>}} \qquad (1.20)$$

Speed $V < c$ for

$$\left|\frac{c(t'_{c_>} - t'_{c_<})}{x'_{c_<} - x'_{c_>}}\right| < 1 \tag{1.21}$$

Let us calculate $t_{c_>} - t_{c_<}$ in the reference frame U

$$t_{c_>} - t_{c_<} = \frac{1}{\sqrt{1 - \frac{V^2}{c^2}}}\left[\frac{V}{c^2}(x'_{c_>} - x'_{c_<}) + (t'_{c_>} - t'_{c_<})\right] =$$

$$\frac{1}{\sqrt{1 - \frac{V^2}{c^2}}}\left[\frac{t'_{c_>} - t'_{c_<}}{x'_{c_>} - x'_{c_<}}(x'_{c_>} - x'_{c_<}) + (t'_{c_>} - t'_{c_<})\right] = 0 \tag{1.22}$$

For the greater $V$ we will have $t_{c_>} - t_{c_<} < 0$. It means that for the spacelike intervals the sign of $t_{c_>} - t_{c_<}$ depends on the speed $V$, i.e. causality relation for spacelike events is not valid.

1.4 Fourier diffusion equation and special relativity theory

The speed of the diffusion signals can be calculated [1]

$$v = \sqrt{2D\omega} \tag{1.23}$$

where

$$D = \frac{\hbar}{m} \tag{1.24}$$

and $\omega$ is the angular frequency of the laser pulses. Considering formula (1.23) and (1.24) one obtains

$$v = c\sqrt{2\frac{\hbar\omega}{mc^2}} \tag{1.25}$$

and $v \geq c$ for $\hbar\omega \geq mc^2$.

From formula (1.25) we conclude that for $\hbar\omega > mc^2$ the Fourier diffusion equation is in contradiction with special relativity theory and breaks the causality in transport phenomena.

It can be shown that the description of the ultrashort thermal energy transport needs the hyperbolic diffusion equation (one dimension transport)

$$\tau \frac{\partial^2 T}{\partial t^2} + \frac{\partial T}{\partial t} = D \frac{\partial^2 T}{\partial x^2} \qquad (1.26)$$

In the equation (1.26) $\tau = \frac{\hbar}{m\alpha^2 c^2}$ is the relaxation time, $m$ = mass of the heat carrier, $a$ is the coupling constant and $c$ is the light speed in vacuum, $T(x,t)$ is the temperature field and $D = \hbar/m$.

Tthe speed of the thermal propagation $v$ can be calculated [1]

$$v = \frac{2\hbar}{m} \sqrt{-\frac{m}{2\hbar}\tau\omega^2 + \frac{m\omega}{2\hbar}(1+\tau^2\omega^2)^{1/2}} \qquad (1.27)$$

Considering that $\tau = \hbar/m\alpha^2 c^2$ formula (1.27) can be written as

$$v = \frac{2\hbar}{m} \sqrt{-\frac{m}{2\hbar}\frac{\hbar\omega^2}{mc^2\alpha^2} + \frac{m\omega}{2\hbar}(1+\frac{\hbar^2\omega^2}{m^2 c^4 \alpha^4})^{1/2}} \qquad (1.28)$$

For

$$\frac{\hbar\omega}{mc^2\alpha^2} < 1, \quad \frac{\hbar\omega^2}{mc^2} < 1 \qquad (1.29)$$

one obtains from formula (1.28)

$$v = \sqrt{\frac{2\hbar}{m}\omega} \qquad (1.30)$$

Formally formula (1.30) is the same as formula (1.25) but considering inequality (1.28) we obtain

$$v = \sqrt{\frac{2\hbar\omega}{m}} = \sqrt{2}\alpha c < c \qquad (1.31)$$

and causality is not broken.

For

$$\frac{\hbar\omega}{mc^2} > 1; \quad \frac{\hbar\omega}{\alpha^2 mc^2} > 1 \qquad (1.32)$$

we obtain from formula (1.28)

$$v = ac, \quad v < c \tag{1.33}$$

Considering formulae (1.31) and (1.33) we conclude that the hyperbolic diffusion equation (1.26) describes the thermal phenomena in accordance with special relativity theory and causality is not broken independently of laser beam energy.

When the amplitude of the laser beam approaches the critical electric field of quantum electrodynamics (Schwinger field) the vacuum becomes polarized and electron – positron pairs are created in vacuum. On a distance equal to the Compton length, $\lambda_C = \hbar/m_e c$, the work of critical field on an electron is equal to the electron rest mass energy $m_e c^2$, i.e. $eE_{Sch}\lambda_C = m_e c^2$. The dimensionless parameter

$$\frac{E}{E_{Sch}} = \frac{e\hbar E}{m_e^2 c^3} \tag{1.34}$$

becomes equal to unity for electromagnetic wave intensity of the order of

$$I = \frac{c}{r_e \lambda_C^2} \frac{m_e c^2}{4\pi} \cong 4.7 \cdot 10^{29} \ \frac{W}{cm^2} \tag{1.35}$$

where $r_e$ is the classical electron radius. For such ultra high intensities the effects of nonlinear quantum electrodynamics plays a key role: laser beams excite virtual electron – positron pairs. As a result the vacuum acquires a finite electric and magnetic susceptibility which lead to the scattering of light by light. The cross section for the photon – photon interaction is given by:

$$\sigma_{\gamma\gamma \to \gamma\gamma} = \frac{973}{10125} \frac{\alpha^3}{\pi^2} r_e^2 \left(\frac{\hbar\omega}{m_e c^2}\right)^6, \tag{1.36}$$

for $\hbar\omega/m_e c^2 < 1$ and reaches its maximum, $\sigma_{max} \approx 10^{-20} cm^2$ for $\hbar\omega \approx m_e c^2$ [3]. Considering formulae (1.35) and (1.36) we conclude that linear hyperbolic diffusion equation is valid only for the laser intensities $I \leq 10^{29}$ W/cm². for high intensities the nonlinear hyperbolic diffusion equation must be formulated and solved.

Table 1.1. Hierarchical structure of the thermal excitation [1]

| Interaction | $a$ | $mc^2 a$ |
|---|---|---|
| Electromagnetic | $137^{-1}$ | $0.5/137$ |
| Strong | $\dfrac{15}{100}$ | $\dfrac{140 \cdot 15}{100}$ for pions |
|  |  | $\dfrac{1000 \cdot 15}{100}$ for nucleons |
| Quark - Quark | 1 | 417* |

* D.H. Perkins, Introduction to high energy physics, Addison – Wesley, USA 1987

# Chapter 2

# Fundamentals of laser pulses interaction with matter

2.1. Hyperbolic versus parabolic heat transport equations

As early as 1956 M. Kac considered a particle moving on line at speed $c$, taking discrete steps of equal size, and undergoing collisions (reversals of direction) at random times, according to a Poisson process of intensity $a$. He showed that the expected position of the particle satisfies either of two difference equations, according to its initial direction. With correct scaling followed by a passage to the limit, the difference equations become a pair of first order partial differential equations (PDE). Differentiating those and adding them yields the hyperbolic diffusion equation

$$\frac{d^2 u}{dt^2} + a\frac{du}{dt} = c\frac{d}{dx}\left(c\frac{du}{dx}\right). \tag{2.1}$$

This is an equation of hyperbolic type. If the lower term (in time) is dropped, it's just the one dimensional wave equation.

R. Hersh proposed the operator generalization of Eq. (2.1):

$$\frac{d}{dt}\left(\frac{dy}{dt}\right) + a\frac{dy}{dt} = A^2 u \tag{2.2}$$

In equation (2.2) $A$ is the generator of a group of linear operators acting on a linear space $B$. Instead of transition moving randomly to the left and right at speed $c$, the time evolution according to generators $A$ and $*A$ is substituted.

The study and applications of the classical hyperbolic diffusion equation (2.1) covers the thermal processes the stock prices, astrophysics and heavy ion physics .

In this paragraph we will study the ultra-short thermal processes in the framework of the hyperbolic diffusion equation.

When an ultrafast thermal pulse (e. g. femtosecond pulse) interacts with a metal surface, the excited electrons become the main carriers of the thermal energy. For a femtosecond thermal pulse, the duration of the pulse is of the same order as the electron relaxation time. In this case, the hyperbolicity of the thermal energy transfer plays an important role.

Radiation deposition of energy in materials is a fundamental phenomenon to laser processing. It converts radiation energy into material's internal energy, which initiates many thermal phenomena, such as heat pulse propagation, melting and evaporation. The operation of many laser techniques requires an accurate understanding and control of the energy deposition and transport processes. Recently, radiation deposition and the subsequent energy transport in metals have been investigated with picosecond and femtosecond resolutions .Results show that during high-power and short-pulse laser heating, free electrons can be heated to an effective temperature much higher than the lattice temperature, which in turn leads to both a much faster energy propagation process and a much smaller lattice-temperature rise than those predicted from the conventional radiation heating model. Corkum et al. found that this electron-lattice nonequilibrium heating mechanism can significantly increase the resistance of molybdenum and copper mirrors to thermal damage during high-power laser irradiation when the laser pulse duration is shorter than one nanosecond. Clemens et al. studied thermal transport in multilayer metals during picosecond laser heating. The measured temperature response in the first 20 ps was found to be different from predictions of the conventional Fourier model. Due to the relatively low temporal resolution of the experiment (~ 4 ps), however, it is difficult to determine whether this difference is the result of nonequilibrium laser heating or is due to other heat conduction mechanisms, such as non-Fourier heat conduction, or reflection and refraction of thermal waves at interfaces. Heat is conducted in solids through electrons and phonons. In metals, electrons dominate the heat conduction, while in insulators and semiconductors, phonons are the major heat carriers. Table 2.1 lists important features of the electrons and phonons.

Table 2.1. General Features of Heat Carriers [1]

|  | Free Electron | Phonon |
|---|---|---|
| Generation | ionization or excitation | lattice vibration |
| Propagation media | vacuum or media | media only |
| Statistics | Fermion | Boson |
| Dispersion | $E = \hbar^2 q^2/(2m)$ | $E = E(q)$ |
| Velocity (m·s$^{-1}$) | ~ $10^6$ | ~ $10^3$ |

The traditional thermal science, or macroscale heat transfer, employs phenomenological laws, such as Fourier's law, without considering the detailed motion of the heat carriers. Decreasing dimensions, however, have brought an increasing need for understanding the heat transfer processes from the microscopic point of view of the heat carriers. The response of the electron and phonon gases to the external perturbation initiated by laser irradiation can be described with the help of a memory function of the system. To that aim, let us consider the generalized Fourier law [1,2]:

$$q(t) = -\int_{-\infty}^{t} K(t-t')\nabla T(t')dt', \qquad (2.3)$$

where $q(t)$ is the density of a thermal energy flux, $T(t')$ is the temperature of electrons and $K(t - t')$ is a memory function for thermal processes. The density of thermal energy flux satisfies the following equation of heat conduction:

$$\frac{\partial}{\partial t}T(t) = \frac{1}{\rho c_v}\nabla^2 \int_{-\infty}^{t} K(t-t')T(t')dt', \qquad (2.4)$$

where $\rho$ is the density of charge carriers and $c_v$ is the specific heat of electrons in a constant volume. We introduce the following equation for the memory function describing the Fermi gas of charge carriers:

$$K(t-t') = K_1 \lim_{t_0 \to 0} \delta(t-t'-t_0). \qquad (2.5)$$

In this case, the electron has a very "short" memory due to thermal disturbances of the state of equilibrium. Combining Eqs. (2.5) and (2.4) we obtain

$$\frac{\partial}{\partial t}T(t) = \frac{1}{\rho c_v} K_1 \nabla^2 T. \tag{2.6}$$

Equation (2.6) has the form of the parabolic equation for heat conduction (PHC). Using this analogy, Eq. (2.6) may be transformed as follows:

$$\frac{\partial}{\partial t}T(t) = D_T \nabla^2 T. \tag{2.7}$$

where the heat diffusion coefficient $D_T$ is defined as follows:

$$D_T = \frac{K_1}{\rho c_v}. \tag{2.8}$$

From Eq. (2.8), we obtain the relation between the memory function and the diffusion coefficient

$$K(t-t') = D_T \rho c_v \lim_{t_0 \to 0} \delta(t-t'-t_0). \tag{2.9}$$

In the case when the electron gas shows a "long" memory due to thermal disturbances, one obtains for memory function

$$K(t-t') = K_2 \tag{2.10}$$

When Eq. (2.10) is substituted to the Eq. (2.4) we obtain

$$\frac{\partial}{\partial t}T = \frac{K_2}{\rho c_v} \nabla^2 \int_{-\infty}^{t} T(t')dt, \tag{2.11}$$

Differentiating both sides of Eq. (2.11) with respect to $t$, we obtain

$$\frac{\partial^2 T}{\partial t^2} = \frac{K_2}{\rho c_v} \nabla^2 T. \tag{2.12}$$

Equation (2.12) is the hyperbolic wave equation describing thermal wave propagation in a charge carrier gas in a metal film. Using a well-known form of the wave equation,

$$\frac{1}{v^2}\frac{\partial^2 T}{\partial t^2} = \nabla^2 T. \tag{2.13}$$

and comparing Eqs. (2.12) and (2.13), we obtain the following form for the memory function:

$$K(t-t') = \rho c_v v^2 \tag{2.14}$$

$v$ = finite, $v < \infty$.

As the third case, "intermediate memory" will be considered:

$$K(t-t') = \frac{K_3}{\tau}\exp\left[-\frac{(t-t')}{\tau}\right], \tag{2.15}$$

where $\tau$ is the relaxation time of thermal processes. Combining Eqs. (2.15) and (2.4) we obtain

$$c_v \frac{\partial^2 T}{\partial t^2} + \frac{c_v}{\tau}\frac{\partial T}{\partial t} = \frac{K_3}{\rho\tau}\nabla^2 T \tag{2.16}$$

and

$$K_3 = D_\tau c_v \rho. \tag{2.17}$$

Thus, finally,

$$\frac{\partial^2 T}{\partial t^2} + \frac{1}{\tau}\frac{\partial T}{\partial t} = \frac{D_\tau}{\tau}\nabla^2 T. \tag{2.18}$$

Equation (2.18) is the hyperbolic equation for heat conduction (HHC), in which the electron gas is treated as a Fermion gas. The diffusion coefficient $D_T$ can be written in the form:

$$D_T = \frac{1}{3}v_F^3 \tau, \tag{2.19}$$

where $v_F$ is the Fermi velocity for the electron gas in a semiconductor. Applying Eq. (2.19) we can transform the hyperbolic equation for heat conduction, Eq. (2.18), as follows:

$$\frac{\partial^2 T}{\partial t^2} + \frac{1}{\tau}\frac{\partial T}{\partial t} = \frac{1}{3}v_F^3 \nabla^2 T. \tag{2.20}$$

Let us denote the velocity of disturbance propagation in the electron gas as $s$:

$$s = \sqrt{\frac{1}{3}}v_F. \tag{2.21}$$

Using the definition of $s$, Eq. (2.20) may be written in the form

$$\frac{1}{s^2}\frac{\partial^2 T}{\partial t^2} + \frac{1}{\tau s^2}\frac{\partial T}{\partial t} = \nabla^2 T. \tag{2.22}$$

For the electron gas, treated as the Fermi gas, the velocity of sound propagation is described by the equation

$$v_s = \left(\frac{P_F^2}{3mm^*}(1+F_0^S)\right)^{1/2}, \qquad P_F = mv_F, \tag{2.23}$$

where $m$ is the mass of a free (non-interacting) electron and $m^*$ is the effective electron mass. Constant $F_0^S$ represents the magnitude of carrier-carrier interaction in the Fermi gas. In the case of a very weak interaction, $m \to m^*$ and $F_0^S \to 0$, so according to Eq. (2.23),

$$v_S = \frac{mv_F}{\sqrt{3m}} = \sqrt{\frac{1}{3}}v_F. \tag{2.24}$$

To sum up, we can make a statement that for the case of weak electron-electron interaction, sound velocity $v_S = \sqrt{1/3}v_F$ and this velocity is equal to the velocity of thermal disturbance propagation $s$. From this we conclude that the hyperbolic equation for heat conduction Eq. (2.22), is identical as the equation for second sound propagation in the electron gas:

$$\frac{1}{v_S^2}\frac{\partial^2 T}{\partial t^2} + \frac{1}{\tau v_S^2}\frac{\partial T}{\partial t} = \nabla^2 T. \tag{2.25}$$

Using the definition expressed by Eq. (2.19) for the heat diffusion coefficient, Eq. (2.25) may be written in the form

$$\frac{1}{v_S^2}\frac{\partial^2 T}{\partial t^2} + \frac{1}{D_T}\frac{\partial T}{\partial t} = \nabla^2 T. \tag{2.26}$$

The mathematical analysis of Eq. (2.25) leads to the following conclusions:

1. In the case when $v_S^2 \to \infty$, $\tau v_S^2$ is finite, Eq. (2.26) transforms into the parabolic equation for heat diffusion:

$$\frac{1}{D_T}\frac{\partial T}{\partial t} = \nabla^2 T. \tag{2.27}$$

2. In the case when $\tau \to \infty$, $v_S$ is finite, Eq. (2.26) transforms into the wave equation:

$$\frac{1}{v_S^2}\frac{\partial^2 T}{\partial t^2} = \nabla^2 T. \tag{2.28}$$

Equation (2.28) describes propagation of the thermal wave in the electron gas. From the point of view of theoretical physics, condition $v_S \to \infty$ violates the special theory of relativity. From this theory we know that there is a limited velocity of interaction propagation and this velocity $v_{lim} = c$, where $c$ is the velocity of light in a vacuum. Multiplying both sides of Eq. (2.26) by $c^2$, we obtain

$$\frac{c^2}{v_S^2}\frac{\partial^2 T}{\partial t^2} + \frac{c^2}{D_T}\frac{\partial T}{\partial t} = c^2 \nabla^2 T, \tag{2.29}$$

Denoting $\beta = v_S/c$, Eq. (2.29) may be written in the form

$$\frac{1}{\beta^2}\frac{\partial^2 T}{\partial t^2} + \frac{1}{\tilde{D}_T}\frac{\partial T}{\partial t} = c^2 \nabla^2 T, \tag{2.30}$$

where $\tilde{D}_T = \tau \beta^2$, $\beta < 1$. On the basis of the above considerations, we conclude that the heat conduction equation, which satisfies the special theory of relativity, acquires the form of the partial hyperbolic Eq. (2.30). The rejection of the first component in Eq. (2.30) violates the special theory of relativity.

Heat transport during fast laser heating of solids has become a very active research area due to the significant applications of short pulse lasers in the fabrication of sophisticated microstructures, synthesis of advanced materials, and measurements of thin film properties. Laser heating of metals involves the deposition of radiation energy on electrons, the energy exchange between electrons, and the lattice, and the propagation of energy through the media.

The theoretical predictions showed that under ultrafast excitation conditions the electrons in a metal can exist out of equilibrium with the lattice for times of the order of the electron energy relaxation time .. Model calculations suggest that it should be possible to heat the electron gas to temperature $T_e$ of up to several thousand degrees for a few picoseconds while keeping the lattice temperature $T_l$ relatively cold. Observing the subsequent equilibration of the electronic system with the lattice allows one to directly study electron-phonon coupling under various conditions.

Several groups have undertaken investigations relating dynamics' changes in the optical constants (reflectivity, transmissivity) to relative changes in electronic temperature. But only recently, the direct measurement of electron temperature has been reported.

The temperature of hot electron gas in a thin gold film ($l$ = 300 Å) was measured, and a reproducible and systematic deviation from a simple Fermi-Dirac (FD) distribution for short time $\Delta t$ ~ 0.4 ps were obtained. The nascent electrons are the electrons created by the direct absorption of the photons prior to any scattering.

Tthe relaxation dynamics of the electron temperature with the hyperbolic heat transport equation (HHT), Eq.(2.26), can be investigated. Conventional laser heating processes which involve a relatively low-energy flux and long laser pulse have been successfully modeled in metal processing and in measuring thermal diffusivity of thin films . However, applicability of these models to short-pulse laser heating is questionable .As it is well known, the Anisimov model does not properly take into account the finite time for the nascent electrons to relax to the FD distribution. In the Anisimov model, the Fourier law for heat diffusion in the electron gas is assumed. However, the diffusion equation is valid only when relaxation time is zero, $\tau$ = 0, and velocity of the thermalization is infinite, $v \to \infty$.

The effects of ultrafast heat transport can be observed in the results of front-pump back probe measurements. The results of these type of experiments can be summarized as follows. Firstly, the measured delays are much shorter than would be expected if heat were carried by the diffusion of electrons in equilibrium with the lattice (tens of picoseconds). This suggests that heat is transported via the electron gas alone, and that the electrons are out of equilibrium with the lattice on this time scale. Secondly, since the delay increases approximately linearly with the sample thickness, the heat transport velocity can be extracted, $v_h \cong 10^8$ cm $\cdot$ s$^{-1}$ = 1$\mu$m $\cdot$ ps$^{-1}$. This is of the same order of magnitude as the Fermi velocity of electrons in gold, 1.4 $\mu$m $\cdot$ ps$^{-1}$.

heat were carried by the diffusion of electrons in equilibrium with the lattice (tens of picoseconds). This suggests, that the heat is transported via the electron gas alone, and that the electrons are out of equilibrium with the lattice on this time scale. Secondly, since the delay increases approximately linearly with the sample thickness, the heat transport velocity can be extracted $v_h \cong 10^8$ cm·s$^{-1}$ = 1$\mu$m·ps$^{-1}$. This is of the same order of magnitude as the Fermi velocity of electrons in Au, 1.4 $\mu$m·ps$^{-1}$.

Since the heat moves at a velocity comparable to $v_F$ - Fermi velocity of the electron gas, it is natural to question exactly how the transport takes place. Since those electrons which lie close to the Fermi surface are the principal contributors to transport, the heat-carrying electrons move at $v_F$. In the limit of lengths longer than the momentum relaxation length, $\lambda$, the random walk behavior is averaged and the electron motion is subject to a diffusion equation. Conversely, on a length scale shorter than $\lambda$, the electrons move ballistically with velocity close to $v_F$. The importance of the ballistic motion may be appreciated by considering the different hot-electron scattering lengths reported in the literature. The electron-electron scattering length in Au, $\lambda_{ee}$ has been calculated]. They find that $\lambda_{ee} \sim (E - E_F)^2$ for electrons close to the Fermi level. For 2-eV electrons $\lambda_{ee} \approx$ 35nm increasing to 80 nm for 1-eV. The electron-phonon scattering length $\lambda_{ep}$ is usually inferred from conductivity data. Using Drude relaxation times, $\lambda_{ep}$ can be computed, $\lambda_{ep} \approx$ 42nm at 273 K. This is shorter than $\lambda_{ee}$, but of the same order of magnitude. Thus, we would expect that both electron-electron and electron-phonon scattering are important on this length scale. However, since conductivity experiments are steady state measurements, the contribution of phonon scattering in a femtosecond regime experiment such as pump-probe ultrafast lasers, is uncertain. In the usual electron-phonon coupling model , one describes the metal as two coupled subsystems, one for electrons and one for phonons. Each subsystem is in local equilibrium so the electrons are characterized by a FD distribution at temperature $T_e$ and the phonon distribution is characterized by a Bose-Einstein distribution at the lattice temperature $T_l$. The coupling between the two systems occurs via the electron-phonon interaction. The time evolution of the energies in the two subsystems is given by the coupled parabolic differential equations (Fourier law). For ultrafast lasers, the duration of pump pulse is of the order of relaxation time in metals . In that case, the parabolic heat conduction equation is not valid and hyperbolic heat transport equation must be used (2.26)

$$\frac{1}{v_S^2}\frac{\partial^2 T}{\partial t^2} + \frac{1}{D_T}\frac{\partial T}{\partial t} = \nabla^2 T, \qquad D_T = \tau v_S^2. \qquad (2.31)$$

In Eq. (2.31), $v_S$ is the thermal wave speed, $\tau$ is the relaxation time and $D_T$ denotes the thermal diffusivity. In the following, Eq. (2.31) will be used to describe the heat transfer in the thin gold films.

To that aim, we define: $T_e$ is the electron gas temperature and $T_l$ is the lattice temperature. The governing equations for nonstationary heat transfer are

$$\frac{\partial T_e}{\partial t} = D_T \nabla^2 T - \frac{D_T}{v_S^2}\frac{\partial^2 T_e}{\partial t^2} - G(T_e - T_l), \qquad \frac{\partial T_l}{\partial t} = G(T_e - T_l). \qquad (2.32)$$

where $D_T$ is the thermal diffusivity, $T_e$ is the electron temperature, $T_e$ is the lattice temperature, and $G$ is the electron-phonon coupling constant. In the following, we will assume that on subpicosecond scale the coupling between electron and lattice is weak and Eq. (2.32) can be replaced by the following equations (2.26):

$$\frac{\partial T_e}{\partial t} = D_T \nabla^2 T - \frac{D_T}{v_S^2}\frac{\partial^2 T_e}{\partial t^2}, \qquad T_l = \text{constant}. \qquad (2.33)$$

Equation (2.33) describes nearly ballistic heat transport in a thin gold film irradiated by an ultrafast ($\Delta t < 1$ ps) laser beam. The solution of Eq. (2.33) for 1D is given by [1]:

$$T(x,t) = \frac{1}{v_S}\int dx' T(x',0)\left[e^{-t/2\tau}\frac{1}{t_0}\Theta(t-t_0) + e^{-t/2\tau}\frac{1}{2\tau}\left\{I_0\left(\frac{(t^2-t_0^2)^{1/2}}{2\tau}\right) + \frac{t}{(t^2-t_0^2)^{1/2}}I_1\left(\frac{(t^2-t_0^2)^{1/2}}{2\tau}\right)\right\}\Theta(t-t_0)\right]$$

(2.34)

where $v_s$ is the velocity of second sound, $t_0 = (x - x')/v_s$ and $I_0$ and $I_1$ are modified Bessel functions and $\Theta(t - t_0)$ denotes the Heaviside function. We are concerned with the solution to Eq. (2.34) for a nearly delta function temperature pulse generated by laser irradiation of the metal surface. The pulse transferred to the surface has the shape:

$$\Delta T_0 = \frac{\beta \rho_E}{C_V v_s \Delta t} \quad \text{for} \quad 0 \leq x \leq v_s \Delta t,$$

$$\Delta T_0 = 0 \quad \text{for} \quad x \geq v_s \Delta t \tag{2.35}$$

In Eq. (2.35), $\rho_E$ denotes the heating pulse fluence, $\beta$ is the efficiency of the absorption of energy in the solid, $C_V(T_e)$ is electronic heat capacity, and $\Delta t$ is duration of the pulse. With $t = 0$ temperature profile described by Eq. (2.35) yields:

$$T(l,t) = \frac{1}{2} \Delta T_0 e^{-t/2\tau} \Theta(t - t_0) \Theta(t_0 + \Delta t - t) \tag{2.36}$$

$$+ \frac{\Delta t}{4\tau} \Delta T_0 e^{-t/2\tau} \left\{ I_0(z) + \frac{t}{2\tau} \frac{1}{z} I_1(z) \right\} \Theta(t - t_0),$$

where $z = (t^2 - t_0^2)^{1/2}/2\tau$ and $t = l/v_s$. The solution to Eq. (2.33), when there are reflecting boundaries, is the superposition of the temperature at $l$ from the original temperature and from image heat source at $\pm 2nl$. This solution is:

$$T(l,t) = \sum_{i=0}^{\infty} \Delta T_0 e^{-t/2\tau} \Theta(t - t_0) \Theta(t_0 + \Delta t - t) + \frac{\Delta t}{4\tau} \Delta T_0 e^{-t/2\tau} \left\{ I_0(z) + \frac{t}{2\tau} \frac{1}{z} I_1(z) \right\} \Theta(t - t_0),$$

$$\tag{2.37}$$

where $t_i = t_0, 3t_0, 5t_0$, $t_0 = l/v_0$. For gold, $C_V(T_e) = C_e(T_e) = \gamma T_e$, $\gamma = 71.5$ Jm$^{-3}$ K$^{-2}$ and Eq. (2.35) yields:

$$\Delta T_0 = \frac{1.4 \times 10^5 \rho_E \beta}{v_s \Delta t T_e} \quad \text{for} \quad 0 \leq x \leq v_s \Delta t$$

$$\Delta T_0 = 0 \quad \text{for} \quad x \geq v_s \Delta t, \tag{2.38}$$

where $\rho_E$ is measured in mJ · cm$^{-2}$, $v_s$ in $\mu$m · ps$^{-1}$, and $\Delta t$ in ps. For $T_e = 300$K :

$$\Delta T_0 = \frac{4.67 \times 10^2 \rho_E \beta}{v_s \Delta t} \quad \text{for} \quad 0 \leq x \leq v_s \Delta t$$

$$\Delta T_0 = 0 \quad \text{for} \quad x \geq v_s \Delta t, \tag{2.39}$$

The model calculations (formulae 2.36 – 2.39) were applied to the description of the experimental results and a fairly good agreement of the theoretical calculations and experimental results was obtained .

2.2 The thermal waves

In the early fifties it was shown by Dingle ,Ward and Wilks and London ,that a density fluctuation in a phonon gas would propagate as a thermal wave - a second sound wave - provided that "losses" from the wave were negligible. In one of their papers, Ward and Wilks indicated they would attempt to look for a second sound wave in sapphire crystals. No results of their experiments were published. Then, for nearly a decade, the subject of "thermal wave" lay dormant. Interest was revived in the sixties, primarily through the efforts of J. A. Krumhansl, R. A. Guyer and C. C. Ackerman. In the paper by Ackerman and Guyer the thermal wave in dielectric solids was experimentally and theoretically investigated. They found a value for the thermal wave velocity in LiF at a very low temperature $T \sim 1$ K, of $v_s \sim 100 - 300$ ms$^{-1}$. In insulators and semiconductors phonons are the major heat carriers. In metals electrons dominate. For long thermal pulses, i.e., when the pulse duration, $\Delta t$, is larger than the relaxation time, $\tau$ , for thermal processes, $\Delta t \gg \tau$, the heat transfer in metals is well described by Fourier diffusion equation. The advent of modern ultrafast lasers opens up the possibility investigating a new mechanism of thermal transport | the thermal wave in an electron gas heated by lasers. The effect of an ultrafast heat transport can be observed in the results of front pump back probe measurements . The results of this type of experiments can be summarized as follows. Firstly, the measured delays are much shorter than it would be expected if the heat were carried by the diffusion of electrons in equilibrium with the lattice (tens of picoseconds). This suggests that the heat is transported via the electron gas alone, and that the electrons are out of equilibrium with the lattice within this time scale. Secondly, since the delay increases approximately linearly with the sample thickness, the heat transport velocity can be determined, $v_h \sim 10^8$ cm s$^{-1}$ = 1$\mu$m ps$^{-1}$. This is of the same order of magnitude as the Fermi velocity of electrons in Au, 1.4 $\mu$m ps$^{-1}$. Kozlowski et al [1] investigated the heat transport in a thin metal film (Au) with the help of the hyperbolic heat conduction equation. It was shown that when the memory of the hot electron gas in metals is taken into account, then the HHT is the dominant equation for heat transfer. The hyperbolic heat conduction equation for heat transfer in an electron gas has the form (2.26)

$$\frac{1}{\left(\frac{1}{3}v_F^2\right)}\frac{\partial^2 T}{\partial t^2} + \frac{1}{\tau\left(\frac{1}{3}v_F^2\right)}\frac{\partial T}{\partial t} = \nabla^2 T. \tag{2.40}$$

If we consider an infinite electron gas, then the Fermi velocity can be calculated

$$v_F \cong bc \tag{2.41}$$

In Eq. (2.41), $c$ is the light velocity in vacuum and $b \sim 10^{-2}$. Considering Eq. (2.41), Eq. (2.40) can be written in a more elegant form:

$$\frac{1}{c^2}\frac{\partial^2 T}{\partial t^2} + \frac{1}{\tau c^2}\frac{\partial T}{\partial t} = \frac{b^2}{3}\nabla^2 T. \tag{2.42}$$

In order to derive the Fourier law from Eq. (2.42), we are forced to break the special theory of relativity and put in Eq. (2.42) $c \to \infty$; $\tau \to 0$. In addition, it can be demonstrated from HHT in a natural way, that in electron gas the heat propagation with velocity $v_h \sim v_F$ in the accordance with the results of the pump probe experiments.

Considering the importance of the thermal wave in future engineering applications and simultaneously the lack of the simple physics presentation of the thermal wave for engineering audience in the following we present the main results concerning the wave nature of heat transfer.

Hence, we discuss Eq. (2.42) in more detail. Firstly, we observe that the second derivative term dominates when:

$$c^2 (\Delta t)^2 < c^2 \Delta t \tau \tag{2.43}$$

i.e., when $\Delta t < \tau$. This implies that for very short heat pulses we have a hyperbolic wave equation of the form:

$$\frac{1}{c^2}\frac{\partial^2 T}{\partial t^2} = \frac{b^2}{3}\nabla^2 T \tag{2.44}$$

and the velocity of the thermal wave is given by

$$v_{th} \sim \frac{1}{\sqrt{3}}\frac{c}{b}, \qquad b \sim 10^{-2}. \tag{2.45}$$

The velocity $v_{th}$ in Eq. (2.45) is the velocity of the thermal wave in an infinite Fermi gas of electrons, which is free of all impurities. The thermal wave, which is described by the solution of Eq. (2.44), does not interact with the crystal lattice. It is the maximum value of the thermal wave obtainable in an infinite free electron gas. If we consider the opposite case to that in Eq. (2.43)

$$c^2(\Delta t)^2 > c^2 \Delta t \tau \tag{2.46}$$

i.e., when

$$\Delta t > \tau \tag{2.47}$$

then, one obtains from Eq. (2.42):

$$\frac{1}{\tau c^2}\frac{\partial T}{\partial t} = \frac{b^2}{3}\nabla^2 T. \tag{2.48}$$

Eq. (2.48) is the parabolic heat conduction equation – Fourier equation.

The solutions of Eq. (2.42) for the following input parameters: $\tau$ = 0.12 ps; $v_{th}$ = 0.15 $\mu$m ps$^{-1}$, $\Delta t$ = 0.02 ps; 0.06 ps; 0.1 ps are presented in Figs. 2.1 – 2.3.

The value of the thermal wave velocity $v_h$ is taken from paper [2.20]. Isotherms are presented as a function of the thin film thickness (length) $l$ [$\mu$m] and the delay times. The mechanism of heat transfer on a nanometer scale, can be divided into three stages: a heat wave for $t \sim Lv_{th}^{-1}$, mixed heat transport for $Lv_{th}^{-1} < t < 3Lv_{th}^{-1}$ and diffusion for $t > 3Lv_{th}^{-1}$. The thermal wave moves in a manner described by the hyperbolic differential partial equation, $x = v_{th} t$. For $t < xv_{th}^{-1}$ the system is undisturbed by an external heat source (laser beam). For longer heat pulses the evidence of the thermal wave is gradually reduced - but the retardation of the thermal pulse is still evident.

If heat is released in a body of gas liquid or solid, a thermal flux transported by heat conduction appears. The pressure gradients associated with the thermal gradients set a gas or liquid in motion, so that additional energy transport occurs through convection. In particular, at sufficiently large energy releases, shock waves are formed in a gas or liquid which transport thermal energy at velocities larger that the speed of sound. Below the critical energy release, nearly pure thermal wave may propagate owing to heat conduction in a gas or liquid with other transport mechanisms being negligible . Solids metals provide an ideal test medium for the study of thermal waves, since they are practically incompressible at temperature below their melting point and the thermal wave pressures are small compared to the classic pressure (produced by repulsion of the atoms in the lattice) up to large energy releases. In accordance with this picture, the speed of sound in a metal is independent of temperature and given by $c_s = (E/\rho)^{1/2}$ where $E$ is the elasticity modulus and $\rho$ is the density.

Using the path-integral method developed in paper [2.31], the solution of the HHT can be obtained. It occurs, that the velocity of the thermal wave in medium is lower than the velocity of the initial thermal wave. The slowing of the thermal wave is caused by the scattering of heat carriers in medium. The scatterings also change the phase of the initial thermal wave.

In one-dimensional flow of heat in metals, the hyperbolic heat transport equation is given by (2.20).

$$\tau \frac{\partial^2 T}{\partial t^2} + \frac{\partial T}{\partial t} = D_T \frac{\partial^2 T}{\partial x^2}, \qquad D_T = \frac{1}{3} v_F^3 \tau, \tag{2.49}$$

where $\tau$ denotes the relaxation time, $D_T$ is the diffusion coefficient and $T$ is the temperature. Introducing the non-dimensional spatial coordinate $z = x/\lambdabar$, where $\lambdabar = \lambda/2\pi$ denotes the reduced mean free path, Eq. (2.49) can be written in the form:

$$\frac{1}{v'^2} \frac{\partial^2 T}{\partial t^2} + \frac{2a}{v'^2} \frac{\partial T}{\partial t} = \frac{\partial^2 T}{\partial z^2}, \tag{2.50}$$

where

$$v' = \frac{v}{\lambda} \qquad a = \frac{1}{2\tau} \tag{2.51}$$

In Eq. (2.51) $v$ denotes the velocity of heat propagation [2.10], $v = (D/\tau)^{1/2}$.

In the paper by C. De Witt-Morette and See Kit Fong the path-integral solution of Eq. (2.50) was obtained. It was shown, that for the initial condition of the form:

$$T(z,0) = \Phi(z) \qquad \text{an "arbitrary" function}$$

$$\frac{\partial T(z,t)}{\partial t}\bigg|_{t=0} = 0 \tag{2.52}$$

the general solution of the Eq. (2.49) has the form:

$$T(z,t) = \frac{1}{2}[\Phi(z,t) + \Phi(z,-t)]e^{-at}$$

$$+ \frac{a}{2}e^{-at}\int_0^t d\eta [\Phi(z,\eta) + \Phi(z,-\eta)] \tag{2.53}$$

$$+ \left[ I_0(a(t^2-\eta^2)^{1/2}) + \frac{t}{(t^2-\eta^2)^{1/2}} I_1(a(t^2-\eta^2)^{1/2}) \right]$$

In Eq. (2.53), $I_0(x)$ and $I_1(x)$ denote the modified Bessel function of zero and first order respectively.

Let us consider the propagation of the initial thermal wave with velocity $v'$, i.e.,

$$\Phi(z - v't) = \sin(z - v't) \tag{2.54}$$

In that case, the integral in (2.53) can be computed analytically, $\Phi(z, t) + \Phi(z, -t) = 2\sin z \cos(v't)$ and the integrals on the right-hand side of (2.53) can be done explicitly [2.31]; we obtain:

$$F(z,t) = e^{-at}\left[\frac{a}{w_1}\sin(w_1 t) + \cos(w_1 t)\right]\sin z, \qquad v' \geq a \tag{2.55}$$

and

$$F(z,t) = e^{-at}\left[\frac{a}{w_2}\sinh(w_2 t) + \cosh(w_2 t)\right]\sin z, \qquad v' < a \tag{2.56}$$

where $w_1 = (v'^2 - a^2)^{1/2}$ and $w_2 = (a^2 - v'^2)^{1/2}$.

In order to clarify the physical meaning of the solutions given by formulas (2.55) and (2.56), we observe that $v' = v/\lambda$ and $w_1$ and $w_2$ can be written as:

$$v_1 = \lambda w_1 = v\left(1 - \left(\frac{1}{2\tau\omega}\right)^2\right)^{1/2}, \qquad 2\tau\omega > 1$$

$$v_2 = \lambda w_2 = v\left(\left(\frac{1}{2\tau\omega}\right)^2 - 1\right)^{1/2}, \qquad 2\tau\omega < 1 \tag{2.57}$$

where $\omega$ denotes the pulsation of the initial thermal wave. From formula (2.57), it can be concluded that we can define the new effective thermal wave velocities $v_1$ and $v_2$. Considering formulas (2.56) and (2.57), we observe that the thermal wave with velocity $v_2$ is very quickly attenuated in time. It occurs that when $\omega^{-1} > 2\tau$, the scatterings of the heat carriers diminish the thermal wave.

It is interesting to observe that in the limit of a very short relaxation time, i.e., when $\tau \to 0$, $v_2 \to \infty$, because for $\tau \to 0$ Eq. (2.49) is the Fourier parabolic equation.

It can be concluded, that for $\omega^{-1} > 2\tau$, the Fourier equation is relevant equation for the description of the thermal phenomena in metals. For $\omega^{-1} > 2\tau$, the scatterings are slower than in the preceding case and attenuation of the thermal wave is weaker. In that case, $\tau \ne 0$ and $v_1$ is always finite:

$$v_1 = v\left(1 - \left(\frac{1}{2\tau\omega}\right)^2\right)^{1/2} < v \tag{2.58}$$

For $\tau \to 0$, i.e., for very rare scatterings $v_1 \to v$ and Eq. (2.49) is a nearly free thermal wave equation. For $\tau$ finite the $v_1 < v$ and thermal wave propagates in the medium with smaller velocity than the velocity of the initial thermal wave.

Considering the formula (2.55), one can define the change of the phase of the initial thermal wave $\beta$, i.e.:

$$\tan[\beta] = \frac{a}{w_1} = \frac{1}{2\tau\omega}\frac{1}{\sqrt{1 - \frac{1}{4\tau^2\omega^2}}}, \qquad 2\tau\omega > 1 \tag{2.59}$$

We conclude that the scatterings produce the change of the phase of the initial thermal wave. For $\tau \to \infty$ (very rare scatterings), $\tan[\beta] = 0$.

2.3 High-order wave equation for thermal transport phenomena

According that the complete Schrödinger equation has the form

$$i\hbar \frac{\partial \Psi}{\partial t} = -\frac{\hbar^2}{2m}\nabla^2 \Psi + V\Psi \qquad (2.60)$$

where $V$ denotes the potential energy, one can obtain the new parabolic quantum heat transport going back to real time $t \to -2it$ and wave function $\Psi \to T$:

$$\frac{\partial T}{\partial t} = \frac{\hbar}{m}\nabla^2 T - \frac{2V}{\hbar}T \qquad (2.61)$$

Equation (2.61) describer the quantum heat transport for $\Delta t > \tau$, where $\tau$ is the relaxation time. For heat transport initiated by ultrashort laser pulses, when $\Delta t > \tau$ one obtains the second order PDE for quantum thermal phenomena

$$\tau \frac{\partial^2 T}{\partial t^2} + \frac{\partial T}{\partial t} = \frac{\hbar}{m}\nabla^2 T - \frac{2V}{\hbar}T \qquad (2.62)$$

Equation (2.62) can be written as

$$\frac{2V\tau}{\hbar}T + \tau\frac{\partial T}{\partial t} + \tau^2 \frac{\partial^2 T}{\partial t^2} = \frac{\tau\hbar}{m}\nabla^2 T. \qquad (2.63)$$

For distortionless thermal phenomena we obtain

$$V\tau \approx \frac{\hbar}{2}. \qquad (2.64)$$

Equation (2.64) is Heisenberg uncertainty relation for thermal quantum phenomena. Substituting equation (2.64) to equation (2.63) we obtain the new form of quantum thermal equation

$$\left(1 + \tau\frac{\partial}{\partial t} + \tau^2 \frac{\partial^2}{\partial t^2}\right)T = \frac{\tau\hbar}{m}\nabla^2 T. \qquad (2.65)$$

It is obvious, from a dimensional analysis, that one can add the fourth term in equation (2.65), i.e.

$$\left(1 + \tau\frac{\partial}{\partial t} + \tau^2 \frac{\partial^2}{\partial t^2} + \tau^3 \frac{\partial^3}{\partial t^3}\right)T = \frac{\tau\hbar}{m}\nabla^2 T. \qquad (2.66)$$

When $V = 0$ equation (2.66) has the form

$$\left(\tau\frac{\partial}{\partial t} + \tau^2\frac{\partial^2}{\partial t^2} + \tau^3\frac{\partial^3}{\partial t^3}\right)T = \frac{\tau\hbar}{m}\nabla^2 T. \tag{2.67}$$

Let us write Eq. (2.67) in the form

$$\kappa\nabla^2 T = \varepsilon\frac{\partial^2 T}{\partial t^2} + \mu\frac{\partial T}{\partial t} + \mu_3\frac{\partial^3 T}{\partial t^3} \tag{2.68}$$

where

$$\kappa = \frac{\tau\hbar}{m}, \qquad \varepsilon = \tau^2, \qquad \mu = \tau, \qquad \mu_3 = \tau^3 \tag{2.69}$$

Equation (2.68) yields the characteristic polynomial equation

$$p(s, jk) = \mu_3 s^3 + \varepsilon s^2 + \mu s + \kappa k^2 = 0 \tag{2.70}$$

Equation (2.68) was investigated, for oscillating transport phenomena, by P.M. Ruiz. In one-dimensional case one obtains from Eq. (2.67)

$$\tau\frac{\partial T}{\partial t} + \tau^2\frac{\partial^2 T}{\partial t^2} + \tau^3\frac{\partial^3 T}{\partial t^3} = \frac{\tau\hbar}{m}\frac{\partial^2 T}{\partial x^2} \tag{2.71}$$

$$\tag{2.72}$$

Below we analyze the third-order wave equation

$$\tau^2\frac{\partial^3 T}{\partial t^3} = \frac{\hbar}{m}\frac{\partial^2 T}{\partial x^2} \tag{2.73}$$

in the case of thermal processes induced by attosecond laser pulses

$$\tau = \frac{\hbar}{mv^2}, \qquad v = \alpha c, \tag{2.74}$$

and equation (2.73) can be rewritten as

$$\frac{\partial^2 T}{\partial x^2} = \beta\frac{\partial^3 T}{\partial t^3}, \qquad \beta = \frac{\hbar}{mv^4}. \tag{2.75}$$

We seek a solution of equation (2.75) of the form

$$T(x,t) = Ae^{i(kx-\omega t)}. \tag{2.76}$$

Substituting equation (2.76) to Eq. (2.75) one obtains

$$(ik)^2 = \beta(-i\omega)^3. \tag{2.77}$$

This shows that equation (2.76) is the solution of the third-order PDE (2.75) i.e. Eq. (2.75) is the third-order wave equation if

$$\beta = \frac{(ik)^2}{(-i\omega)^3} = \frac{\hbar}{mv^2} \qquad (2.78)$$

where $v$ is the speed of propagation of thermal energy [2.33]. Substituting Eq. (2.78) to Eq. (2.76) one obtains

$$T(x,t) = Ae^{i\left[\frac{x}{\sqrt{2}\lambda}-\omega t\right]}e^{-\frac{1}{\sqrt{2}}\frac{x}{\lambda}} + Be^{-i\left[\frac{x}{\sqrt{2}\lambda}-\omega t\right]}e^{\frac{1}{\sqrt{2}}\frac{x}{\lambda}} \qquad (2.79)$$

where $\lambda$ is mean free path.

The second term in Eq. (2.79) tends to infinity for $x/\lambda \gg 1$ and is to be omitted. The final solution of Eq. (2.76) has the form

$$T(x,t) = e^{-\frac{1}{\sqrt{2}}\frac{x}{\lambda}}Ae^{i\left[\frac{x}{\sqrt{2}\lambda}-\omega t\right]} \qquad (2.80)$$

and describes the strongly damped thermal wave.

It is interesting to observe that for electromagnetic interaction the third-order time derivative $d^3x/dt^3$ also describes the damping of the electron motion due to the self interaction of the charges .

# Chapter 3

# Quantum heat transport equation

3.1 Quantum parabolic and hyperbolic heat transport equations

Dynamical processes are commonly investigated using laser pump-probe experiments with a pump pulse exciting the system of interest and a second probe pulse tracking is temporal evolution. As the time resolution attainable in such experiments depends on the temporal definition of the laser pulse, pulse compression to the attosecond domain is a recent promising development.

After the standards of time and space were defined the laws of classical physics relating such parameters as distance, time, velocity, temperature are assumed to be independent of accuracy with which these parameters can be measured. It should be noted that this assumption does not enter explicitly into the formulation of classical physics. It implies that together with the assumption of existence of an object and really independently of any measurements (in classical physics) it was tacitly assumed that *there was a possibility of an unlimited increase in accuracy of measurements.* Bearing in mind the "atomicity" of time i.e. considering the smallest time period, the Planck time, the above statement is obviously not true. Attosecond laser pulses we are at the limit of laser time resolution.

With attosecond laser pulses belong to a new Nano – World where size becomes comparable to atomic dimensions, where transport phenomena follow different laws from that in the macro world. This first stage of miniaturization, from $10^{-3}$ m to $10^{-6}$ m is over and the new one, from $10^{-6}$ m to $10^{-9}$ m just beginning. The Nano – World is a quantum world with all the predicable and non-predicable (yet) features.

In this paragraph, we develop and solve the quantum relativistic heat transport equation for Nano – World transport phenomena where external forces exist.

There is an impressive amount of literature on hyperbolic heat transport in matter. In Chapter 2 we developed the new hyperbolic heat transport equation which generalizes the Fourier heat transport equation for the rapid thermal processes. The hyperbolic heat transport equation (HHT) for the fermionic system has be written in the form (2.25)

$$\frac{1}{\left(\frac{1}{3}v_F^2\right)}\frac{\partial^2 T}{\partial t^2} + \frac{1}{\tau\left(\frac{1}{3}v_F^2\right)}\frac{\partial T}{\partial t} = \nabla^2 T , \qquad (3.1)$$

where $T$ denotes the temperature, $\tau$ the relaxation time for the thermal disturbance of the fermionic system, and $v_F$ is the Fermi velocity.

In what follows we develop the new formulation of the HHT, considering the details of the two fermionic systems: electron gas in metals and the nucleon gas.

For the electron gas in metals, the Fermi energy has the form [1]

$$E_F^e = (3\pi)^2 \frac{n^{2/3}\hbar^2}{2m_e}, \qquad (3.2)$$

where $n$ denotes the density and $m_e$ electron mass. Considering that

$$n^{-1/3} \sim a_B \sim \frac{\hbar^2}{me^2}, \qquad (3.3)$$

and $a_B$ = Bohr radius, one obtains

$$E_F^e \sim \frac{n^{2/3}\hbar^2}{2m_e} \sim \frac{\hbar^2}{ma^2} \sim \alpha^2 m_e c^2, \qquad (3.4)$$

where $c$ = light velocity and $\alpha = 1/137$ is the fine-structure constant for electromagnetic interaction. For the Fermi momentum $p_F$ we have

$$p_F^e \sim \frac{\hbar}{a_B} \sim \alpha m_e c, \qquad (3.5)$$

and, for Fermi velocity $v_F$,

$$v_F^e \sim \frac{p_F}{m_e} \sim \alpha c. \qquad (3.6)$$

Formula (3.6) gives the theoretical background for the result presented in Chapter 2. Comparing formulas (2.41) and (3.6) it occurs that $b = a$. Considering formula (3.6), Eq. (2.42) can be written as

$$\frac{1}{c^2}\frac{\partial^2 T}{\partial t^2} + \frac{1}{c^2\tau}\frac{\partial T}{\partial t} = \frac{\alpha^2}{3}\nabla^2 T. \tag{3.7}$$

As seen from (3.7), the HHT equation is a relativistic equation, since it takes into account the finite velocity of light.

For the nucleon gas, Fermi energy equals

$$E_F^N = \frac{(9\pi)^{2/3}\hbar^2}{8mr_0^2}, \tag{3.8}$$

where $m$ denotes the nucleon mass and $r_0$, which describes the range of strong interaction, is given by

$$r_0 = \frac{\hbar}{m_\pi c}, \tag{3.9}$$

wherein $m_\pi$ is the pion mass. From formula (3.9), one obtains for the nucleon Fermi energy

$$E_F^N \sim \left(\frac{m_\pi}{m}\right)^2 mc^2. \tag{3.10}$$

In analogy to the Eq. (3.4), formula (3.10) can be written as

$$E_F^N \sim \alpha_s^2 mc^2, \tag{3.11}$$

where $\alpha_s = \frac{m_\pi}{m} \cong 0.15$ is the fine-structure constant for strong interactions. Analogously, we obtain the nucleon Fermi momentum

$$p_F^e \sim \frac{\hbar}{r_0} \sim \alpha_s mc \tag{3.12}$$

and the nucleon Fermi velocity

$$v_F^N \sim \frac{pF}{m} \sim \alpha_s c, \tag{3.13}$$

and HHT for nucleon gas can be written as

$$\frac{1}{c^2}\frac{\partial^2 T}{\partial t^2}+\frac{1}{c^2\tau}\frac{\partial T}{\partial t}=\frac{\alpha_s^2}{3}\nabla^2 T. \tag{3.14}$$

In the following, the procedure for the discretization of temperature $T(\vec{r},t)$ in hot fermion gas will be developed. First of all, we introduce the reduced de Broglie wavelength

$$\lambda_B^e=\frac{\hbar}{m_e v_h^e}, \qquad v_h^e=\frac{1}{\sqrt{3}}\alpha c,$$
$$\lambda_B^N=\frac{\hbar}{m v_h^N}, \qquad v_h^N=\frac{1}{\sqrt{3}}\alpha_s c, \tag{3.15}$$

and the mean free paths $\lambda_e$ and $\lambda_N$

$$\lambda^e=v_h^e \tau^e, \qquad \lambda^N=v_h^N \tau^N. \tag{3.16}$$

In view of formulas (3.15) and (3.16), we obtain the HHC for electron and nucleon gases

$$\frac{\lambda_B^e}{v_h^e}\frac{\partial^2 T}{\partial t^2}+\frac{\lambda_B^e}{\lambda^e}\frac{\partial T}{\partial t}=\frac{\hbar}{m_e}\nabla^2 T^e, \tag{3.17}$$

$$\frac{\lambda_B^N}{v_h^N}\frac{\partial^2 T}{\partial t^2}+\frac{\lambda_B^N}{\lambda^N}\frac{\partial T}{\partial t}=\frac{\hbar}{m}\nabla^2 T^N. \tag{3.18}$$

Equations (3.17) and (3.18) are the hyperbolic partial differential equations which are the master equations for heat propagation in Fermi electron and nucleon gases. In the following, we will study the quantum limit of heat transport in the fermionic systems. We define the quantum heat transport limit as follows:

$$\lambda^e=\lambdabar_B^e, \qquad \lambda^N=\lambdabar_B^N. \tag{3.19}$$

In that case, Eqs. (3.17) and (3.18) have the form

$$\tau^e \frac{\partial^2 T^e}{\partial t^2}+\frac{\partial T^e}{\partial t}=\frac{\hbar}{m_e}\nabla^2 T^e, \tag{3.20}$$

$$\tau^N \frac{\partial^2 T^N}{\partial t^2}+\frac{\partial T^N}{\partial t}=\frac{\hbar}{m}\nabla^2 T^N, \tag{3.21}$$

where

$$\tau^e=\frac{\hbar}{m_e (v_h^e)^2}, \qquad \tau^N=\frac{\hbar}{m (v_h^N)^2}. \tag{3.22}$$

Equations (3.20) and (3.21) define the master equation for quantum heat transport (QHT). Having the relaxation times $\tau^e$ and $\tau^N$, one can define the "pulsations" $\omega_h^e$ and $\omega_h^N$

$$\omega_h^e = (\tau^e)^{-1}, \qquad \omega_h^N = (\tau^N)^{-1}, \tag{3.23}$$

or

$$\omega_h^e = \frac{m_e (v_h^e)^2}{\hbar}, \qquad \omega_h^N = \frac{m (v_h^N)^2}{\hbar},$$

i.e.,

$$\begin{aligned}\omega_h^e \hbar &= m_e (v_h^e)^2 = \frac{m_e \alpha^2}{3} c^2, \\ \omega_h^N \hbar &= m (v_h^N)^2 = \frac{m \alpha_s^2}{3} c^2.\end{aligned} \tag{3.24}$$

The formulas (3.24) define the Planck-Einstein relation for heat quanta $E_h^e$ and $E_h^N$

$$\begin{aligned}E_h^e &= \omega_h^e \hbar = m_e (v_h^e)^2, \\ E_h^N &= \omega_h^N \hbar = m_N (v_h^N)^2.\end{aligned} \tag{3.25}$$

The heat quantum with energy $E_h = \hbar \omega$ can be named the *heaton*, in complete analogy to the *phonon, magnon, roton*, etc. For $\tau^e, \tau^N \to 0$, Eqs. (3.20) and (3.24) are the Fourier equations with quantum diffusion coefficients $D^e$ and $D^N$

$$\frac{\partial T^e}{\partial t} = D^e \nabla^2 T^e, \qquad D^e = \frac{\hbar}{m_e}, \tag{3.26}$$

$$\frac{\partial T^N}{\partial t} = D^N \nabla^2 T^N, \qquad D^N = \frac{\hbar}{m}. \tag{3.27}$$

The quantum diffusion coefficients $D^e$ and $D^N$ were introduced for the first time by E. Nelson .

For finite $\tau^e$ and $\tau^N$, for $\Delta t < \tau^e$, $\Delta t < \tau^N$, Eqs. (3.20) and (3.21) can be written as

$$\frac{1}{(v_h^e)^2} \frac{\partial^2 T^e}{\partial t^2} = \nabla^2 T^e, \tag{3.28}$$

$$\frac{1}{(v_h^N)^2}\frac{\partial^2 T^N}{\partial t^2} = \nabla^2 T^N. \tag{3.29}$$

Equations (3.28) and (3.29) are the wave equations for quantum heat transport (QHT). For Δt >τ, one obtains the Fourier equations (3.26) and (3.27).

In what follows, the dimensionless form of the QHT will be used. Introducing the reduced time $t'$ and reduced length $x'$,

$$t' = t/\tau, \qquad x' = \frac{x}{v_h \tau}, \tag{3.30}$$

one obtains, for QHT,

$$\frac{\partial^2 T^e}{\partial t^2} + \frac{\partial T^e}{\partial t} = \nabla^2 T^e, \tag{3.31}$$

$$\frac{\partial^2 T^N}{\partial t^2} + \frac{\partial T^N}{\partial t} = \nabla^2 T^N. \tag{3.32}$$

and, for QFT,

$$\frac{\partial T^e}{\partial t} = \nabla^2 T^e, \tag{3.33}$$

$$\frac{\partial T^N}{\partial t} = \nabla^2 T^N. \tag{3.34}$$

3.2 Proca thermal equation

Electromagnetic phenomena in vacuum are characterized by two three dimensional vector fields, the electric and magnetic fields $E(x,t)$ and $B(x,t)$ which are subject to Maxwell's equation and which can also be thought of as the classical limit of the quantum mechanical description in terms of photons. The photon mass is ordinarily assumed to be exactly zero in Maxwell's electromagnetic field theory, which is based on gauge invariance. If gauge invariance is abandoned, a mass term can be added to the Lagrangian density for the electromagnetic field in a unique way:

$$L = -\frac{1}{4\mu_0} F_{\mu\nu} F^{\mu\nu} - j_\mu A_\mu + \frac{\mu_\gamma^2}{2\mu_0} A_\mu A^\mu \qquad (3.35)$$

where $\mu_\gamma^{-1}$ is a characteristic length associated with photon rest mass, $A_\mu$ and $j_\mu$ are the four-dimensional vector potential $(\vec{A}, i\phi/c)$ and four-dimensional vector current density $(\vec{J}, ic_\rho)$ with $\phi$ and $\vec{A}$ denoting the scalar and vector potential and $\rho$, $\vec{J}$ are the charge and current density, respectively, $\mu_0$ is the permeability constant of free space and $F_{\mu\nu}$ is the antisymmetric field strength tensor. It is connected to the vector potential through

$$F_{\mu\nu} = \frac{\partial A_\nu}{\partial x_\mu} - \frac{\partial A_\mu}{\partial x_\nu} \qquad (3.36)$$

The variation of Lagrangian density with respect to $A_\mu$ yields the Proca equation (Proca 1930)

$$\frac{\partial F_{\mu\nu}}{\partial x_\nu} + \mu_\gamma^2 A_\mu = \mu_0 J_\mu \qquad (3.37)$$

Substituting Eq.(3.36) into (3.37) we obtain the wave equation of the Proca field

$$\left( \nabla^2 - \frac{\partial^2}{\partial(ct)^2} - \mu_\gamma^2 \right) A_\mu = -\mu_0 J \qquad (3.38)$$

$$\left( \Box - \mu_\gamma^2 \right) A_\mu = -\mu_0 J \qquad (3.39)$$

In free space The *Proca* equation (3.39) reduces to (3.40), for a vector electromagnetic potential of $A_\mu$.

$$\left( \Box + \mu_\gamma^2 \right) A_\mu = 0,$$
$$\Box = \frac{1}{c^2} \frac{\partial^2}{\partial t^2} - \frac{\partial^2}{\partial x^2} \qquad (3.40)$$

which is essentially the Klein – Gordon equation for massive photons. The parameter $\mu_\gamma$ can be interpreted as the photon rest mass $m_\gamma$ with

$$m_\gamma = \frac{\mu_\gamma \hbar}{c}. \qquad (3.41)$$

With this interpretation the characteristic scaling length $\mu_\gamma^{-1}$ becomes the reduced Compton wavelength of the photon interaction. An additional point is that static electric and magnetic fields would exhibit exponential dumping governed by the term $\exp(-\mu_\gamma^{-1} r)$ is the photon is massive instead of massless.

3.3 Special relativity with nonzero photon mass

It is well-known that the electromagnetic constant $c$ in Maxwell theory of electromagnetic waves propagating in vacuum and special relativity was developed as a consequence of the constancy of the speed of light. However, one of the prediction of massive photon electromagnetic theory is that there will be dispersion of the velocity of massive photon in vacuum.

The plane wave solution of the Proca equations without current is $A^\nu \sim \exp(ik^\mu x_\mu)$, where the wave vector $k^\mu = (\omega, \vec{k})$ satisfies the relationship

$$k^2 c^2 = \omega^2 - \mu_\gamma^2 c^2 \qquad (3.42)$$

As can be shown that

$$v_g \text{ (group velocity)} = c\left(1 - \frac{\mu_\gamma^2 c^2}{2\omega^2}\right) \qquad (3.43)$$

and

$$v_g = 0 \quad \text{for} \quad \omega = \mu_\gamma c$$

namely the massive waves do not propagate. When $\omega < \mu_\gamma c$, $k$ becomes an imaginary quantity and the amplitude of a free massive wave would, therefore, be attenuated exponentially. Only when $\omega > \mu_\gamma c$ can the waves propagate in vacuum unattenuated. In the limit $\omega \to \infty$, the group velocity will approach the constant $c$ for all phenomena.

A nonzero photon mass implies that the speed of light is not unique constant but is a function of frequency. In fact, the assumption of the constancy of speed of light is not necessary for the validity of special relativity (Szymacha), i.e. special relativity can instead be based on the existence of a unique limiting speed $c$ to which speeds of all bodies tend when their energy becomes much larger then their mass. Then, the velocity that enters in the Lorentz transformation would simply be this limiting speed, not the speed of light

It is quite interesting that the *Proca* type equation can be obtained for thermal phenomena. In the following starting with the hyperbolic heat diffusion equation the *Proca* equation for thermal processes will be developed and solved.

The relativistic hyperbolic transport equation can be written as

$$\frac{1}{v^2}\frac{\partial^2 T}{\partial t^2} + \frac{m_0 \gamma}{\hbar}\frac{\partial T}{\partial t} = \nabla^2 T. \qquad (3.44)$$

In equation (3.44) $v$ is the velocity of heat waves, $m_0$ is the mass of heat carrier and $\gamma$ – the Lorentz factor, $\gamma = \left(1-\frac{v^2}{c^2}\right)^{-\frac{1}{2}}$. As was shown in paper [3.1] the heat energy (*heaton temperature*) $T_h$ can be defined as follows:

$$T_h = m_0 \gamma v^2. \qquad (3.45)$$

Considering that $v$, the thermal wave velocity equals :

$$v = ac \qquad (3.46)$$

where $a$ is the coupling constant for the interactions which generate the *thermal wave* ($a = 1/137$ and $a = 0.15$ for electromagnetic and strong forces respectively). The *heaton temperature* is equal to

$$T_h = \frac{m_0 \alpha^2 c^2}{\sqrt{1-\alpha^2}}. \qquad (3.47)$$

Based on equation (3.47) one concludes that the *heaton temperature* is a linear function of the mass $m_0$ of the heat carrier. It is interesting to observe that the proportionality of $T_h$ and the heat carrier mass $m_0$ was observed for the first time in ultrahigh energy heavy ion reactions measured at CERN   M. Kozlowski et al    [1] has  shown that the temperature of pions, kaons and protons produced in Pb+Pb, S+S reactions are proportional to the mass of particles. Recently, at Rutherford Appleton Laboratory (RAL), the VULCAN    LASER was used to produce the elementary particles: electrons and pions .

In this chapter the forced relativistic heat transport equation will be studied and solved. M. Kozlowski et al developed thermal wave equation

$$\frac{1}{v^2}\frac{\partial^2 T}{\partial t^2} + \frac{m}{\hbar}\frac{\partial T}{\partial t} + \frac{2Vm}{\hbar^2}T - \nabla^2 T = 0. \tag{3.48}$$

The relativistic generalization of equation (3.48) is quite obvious:

$$\frac{1}{v^2}\frac{\partial^2 T}{\partial t^2} + \frac{m_0\gamma}{\hbar}\frac{\partial T}{\partial t} + \frac{2Vm_0\gamma}{\hbar^2}T - \nabla^2 T = 0. \tag{3.49}$$

It is worthwhile noting that in order to obtain a non-relativistic equation we put $\gamma = 1$.

When the external force is present $F(x,t)$ the forced damped heat transport is obtained instead of equation (3.49) (in one dimensional case):

$$\frac{1}{v^2}\frac{\partial^2 T}{\partial t^2} + \frac{m_0\gamma}{\hbar}\frac{\partial T}{\partial t} + \frac{2Vm_0\gamma}{\hbar^2}T - \frac{\partial^2 T}{\partial x^2} = F(x,t). \tag{3.50}$$

The hyperbolic relativistic quantum heat transport equation, (3.50), describes the forced motion of heat carriers which undergo scattering ($\frac{m_0\gamma}{\hbar}\frac{\partial T}{\partial t}$ term) and are influenced by the potential term ($\frac{2Vm_0\gamma}{\hbar^2}T$).

Equation (3.50) can be written as

$$\left(\Diamond + \frac{2Vm_0\gamma}{\hbar^2}\right)T + \frac{m_0\gamma}{\hbar}\frac{\partial T}{\partial t} = F(x,t),$$

$$\Diamond = \frac{1}{v^2}\frac{\partial^2}{\partial t^2} - \frac{\partial^2}{\partial x^2}. \tag{3.51}$$

We seek the solution of equation (3.51) in the form

$$T(x,t) = e^{-\frac{1}{2\tau}t}u(x,t) \tag{3.52}$$

where $\tau = \hbar/(mv^2)$ is the relaxation time. After substituting equation (3.52) in equation (3.51) we obtain a new equation

$$(\Diamond + q^2)u(x,t) = e^{\frac{1}{2\tau}t}F(x,t) \tag{3.53}$$

and

$$q^2 = \frac{2Vm}{\hbar^2} - \left(\frac{mv}{2\hbar}\right)^2 \tag{3.54}$$

$$m = m_0 \gamma \tag{3.55}$$

In free space i.e. when $F(x,t) \to 0$ equation (3.53) reduces to

$$(\lozenge + q^2)u(x,t) = 0 \tag{3.56}$$

which is essentially the free *Proca* equation, compare equation (3.40).

The *Proca* equation describes the interaction of the laser pulse with the matter. As was shown by Kozlowski et al., the quantization of the temperature field leads to the *heatons* – quanta of thermal energy with a mass $m_h = \hbar/{\tau v_h^2}$ ,where $\tau$ is the relaxation time and $v_h$ is the finite velocity for heat propagation. For $v_h \to \infty$, i.e. for $c \to \infty$, $m_o \to 0$. it can be concluded that in non-relativistic approximation ($c$ = infinite) the *Proca* equation is the diffusion equation for massless photons and heatons.

3.4 . Solution of the *Proca* thermal equation

For the initial *Cauchy* condition:

$$u(x,0) = f(x), \qquad u_t(x,0) = g(x) \tag{3.57}$$

the solution of the *Proca* equation has the form (for $q > 0$) [3]

$$\begin{aligned} u(x,t) =\ & \frac{f(x-vt) + f(x+vt)}{2} \\ & + \frac{1}{2v} \int_{x-vt}^{x+vt} g(\varsigma) J_0\left[q\sqrt{v^2 t^2 - (x-\varsigma)^2}\right] d\varsigma \\ & - \frac{\sqrt{q}vt}{2} \int_{x-vt}^{x+vt} f(\varsigma) \frac{J_1\left[q\sqrt{v^2 t^2 - (x-\varsigma)^2}\right]}{\sqrt{v^2 t^2 - (x-\varsigma)^2}} d\varsigma \\ & + \frac{1}{2v} \int_0^t \int_{x-v(t-t')}^{x+v(t-t')} G(\varsigma,t') J_0\left[q\sqrt{v^2 (t-t')^2 - (x-\varsigma)^2}\right] dt' d\varsigma. \end{aligned} \tag{3.58}$$

where $G = e^{1/2\tau} F(x,t)$.

When $q < 0$ solution of *Proca* equation has the form [1,3]:

$$u(x,t) = \frac{f(x-vt) + f(x+vt)}{2}$$

$$+ \frac{1}{2v} \int_{x-vt}^{x+vt} g(\varsigma) I_0\left[-q\sqrt{v^2 t^2 - (x-\varsigma)^2}\right] d\varsigma$$

$$- \frac{\sqrt{-q}vt}{2} \int_{x-vt}^{x+vt} f(\varsigma) \frac{I_1\left[-q\sqrt{v^2 t^2 - (x-\varsigma)^2}\right]}{\sqrt{v^2 t^2 - (x-\varsigma)^2}} d\varsigma \qquad (3.59)$$

$$+ \frac{1}{2v} \int_0^t \int_{x-v(t-t')}^{x+v(t-t')} G(\varsigma,t') I_0\left[-q\sqrt{v^2 (t-t')^2 - (x-\varsigma)^2}\right] dt' d\varsigma.$$

When $q = 0$ equation (3.53) is the forced thermal equation

$$\frac{1}{v^2} \frac{\partial^2 u}{\partial t^2} - \frac{\partial^2 u}{\partial x^2} = G(x,t). \qquad (3.60)$$

On the other hand one can say that equation (3.60) is distortion-less hyperbolic equation. The condition $q = 0$ can be rewritten as:

$$V\tau = \frac{\hbar}{8} \qquad (3.61)$$

The equation (3.61) is the analogous to the Heisenberg uncertainty relation. Considering equation (3.45) equation (3.61) can be written as:

$$V = \frac{T_h}{8}, \qquad V < T_h. \qquad (3.62)$$

It can be stated that distortion-less waves can be generated only if $T_h > V$. For $T_h < V$, i.e. when the "Heisenberg rule" is broken, the shape of the thermal waves is changed.

# Chapter 4
# Heat transport in nanoscale

## 4.1 Pauli-Heisenberg inequality

Clusters and aggregates of atoms in the nanometer range (currently called nanoparticles) are systems intermediate in several respects, between simple molecules and bulk materials and have been the subject of intensive work.

In this paragraph, we investigate the thermal relaxation phenomena in nanoparticles – microtubules within the frame of the quantum heat transport equation. In reference [4.1], the thermal inertia of materials, heated with laser pulses faster than the characteristic relaxation time was investigated. It was shown, that in the case of the ultra-short laser pulses it was necessary to use the hyperbolic heat conduction (HHC). For microtubules the diameters are of the order of the de Broglie wave length. In that case quantum heat transport must be used to describe the transport phenomena,

$$\tau \frac{\partial^2 T}{\partial t^2} + \frac{\partial T}{\partial t} = \frac{\hbar}{m} \nabla^2 T, \qquad (4.1)$$

where $T$ denotes the temperature of the heat carrier, and $m$ denotes its mass and $\tau$ is the relaxation time. The relaxation time $\tau$ is defined as:

$$\tau = \frac{\hbar}{m v_h^2}, \qquad (4.2)$$

where $v_h$ is the thermal pulse propagation rate

$$v_h = \frac{1}{\sqrt{3}} \alpha c \qquad (4.3)$$

In equation (4.3) $\alpha$ is a coupling constant (for the electromagnetic interaction $\alpha = e^2/\hbar c$ and $c$ denotes the speed of light in vacuum. Both parameters $\tau$ and $v_h$ characterizes completely the thermal energy transport on the atomic scale and can be termed *"atomic relaxation time"* and *"atomic"* heat diffusivity.

Both τ and $v_h$ contain constants of Nature, $a$, $c$. Moreover, on an atomic scale there is no shorter time period than and smaller velocity than that build from of constants in Nature. Consequently, one can call τ and $v_h$ the *elementary relaxation time* and *elementary diffusivity,* which characterizes heat transport in the elementary building block of matter, the atom. In the following, starting with elementary τ and $v_h$, we shall describe thermal relaxation processes in microtubules which consist of the $N$ components (molecules) each with elementary τ and $v_h$. With this in view, we use the Pauli-Heisenberg inequality [4.1]

$$\Delta r \Delta p \geq N^{\frac{1}{3}} \hbar, \qquad (4.4)$$

where $r$ denotes the characteristic dimension of the nanoparticle and $p$ is the momentum of the energy carriers. The Pauli-Heisenberg inequality expresses the basic property of the $N$ – fermionic system. In fact, compared to the standard Heisenberg inequality

$$\Delta r \Delta p \geq \hbar, \qquad (4.5)$$

we observe that, in this case the presence of the large number of identical fermions forces the system either to become spatially more extended for a fixed typical momentum dispension, or to increase its typical momentum dispension for a fixed typical spatial extension. We could also say that for a fermionic system in its ground state, the average energy per particle increases with the density of the system.

An illustrative means of interpreting the Pauli-Heisenberg inequality is to compare Eq. (4.4) with Eq. (4.5) and to think of the quantity on the right hand side of it as the *effective fermionic Planck constant*

$$\hbar^f(N) = N^{\frac{1}{3}} \hbar. \qquad (4.6)$$

We could also say that antisymmetrization, which typifies fermionic amplitudes amplifies those quantum effects which are affected by the Heisenberg inequality.

Based on equation (4.6), we can recalculate the relaxation time *τ*, equation (4.2) and the thermal speed $v_h$, equation (4.3) for a nanoparticle consisting of *N* fermions

$$\hbar \leftarrow \hbar^f(N) = N^{\frac{1}{3}}\hbar \qquad (4.7)$$

and obtain

$$v_h^f = \frac{e^2}{\hbar^f(N)} = \frac{1}{N^{\frac{1}{3}}}v_h, \qquad (4.8)$$

$$\tau^f = \frac{\hbar^f}{m(v_h^f)^2} = N\tau. \qquad (4.9)$$

The number $N$ particles in a nanoparticle (sphere with radius $r$) can be calculated using the equation (we assume that density of a nanoparticle does not differ too much from that of the bulk material)

$$N = \frac{\frac{4\pi}{3}r^3 \rho A Z}{\mu} \qquad (4.10)$$

and for non spherical shapes with semi axes $a, b, c$

$$N = \frac{\frac{4\pi}{3}abc \rho A Z}{\mu} \qquad (4.11)$$

where $\rho$ is the density of the nanoparticle, $A$ is the Avogardo number, $\mu$ is the molecular mass of theparticles in grams and $Z$ is the number of valence electrons.

Using equations (4.8) and (4.9), we can calculate the de Broglie wave length $\lambda_B^f$ and mean free path $\lambda_{mfp}^f$ for nanoparticles

$$\lambda_B^f = \frac{\hbar^f}{mv_{th}^f} = N^{\frac{2}{3}}\lambda_B, \qquad (4.12)$$

$$\lambda_{emfp}^f = v_{th}^f \tau_{th}^f = N^{\frac{2}{3}}\lambda_{mfp}, \qquad (4.13)$$

where $\lambda_B$ and $\lambda_{mfp}$ denote the de Broglie wave length and the mean free path for heat carriers in nanoparticles (e.g. microtubules).Microtubules are essential to cell functions. In neurons, microtubules help and regulate synaptic activity responsible for learning and cognitive function. Whereas microtubules have traditionally been considered to be purely structural elements, recent evidence has revealed that mechanical, chemical and electrical signaling and a communication function also exist as a result of the microtubule interaction with membrane structures by linking proteins, ions and voltage fields respectively. The characteristic dimensions of the microtubules; a crystalline cylinder 10 nm internal diameter, are of the order of the de Broglie length for electrons in atoms. When the characteristic length of the structure is of the order of the de Broglie wave length, then the signaling phenomena must be described by the quantum transport theory. In order to describe quantum transport phenomena in microtubules it is necessary to use equation (1) with the relaxation time described by equation

$$\tau = \frac{2\hbar}{mv^2} = \frac{\hbar}{E}. \tag{4.14}$$

The relaxation time is the de-coherence time, i.e. the time before the wave function collapses, when the transition classical $\rightarrow$ quantum phenomena is considered.

In the following we consider the time $\tau$ for atomic and multiatomic phenomena.

$$\tau_a \approx 10^{-17}\,\text{s} \tag{4.15}$$

and when we consider multiatomic transport phenomena, with $N$ equal number of aggregates involved the equation is (4.9)

$$\tau_N = N\tau_a \tag{4.16}$$

4.2 The Penrose-Hameroff model

The Penrose – Hameroff Orchestrated Objective Reduction Model (OrchOR) [4.2] proposes that quantum superposition – computation occurs in nanotubule automata within brain neurons and glia. Tubulin subunits within microtubules act as qubits, switching between states on a nanosecond scale, governed by London forces in hydrophobic pockets. These oscillations are tuned and orchestrated by microtubule associated proteins (MAPs) providing a feedback loop between the biological system and the quantum state. These qubits interact computationally by non-localquantum entanglement, according to the Schrödinger equation with preconscious processing continuing until the threshold for objective reduction (OR) is reached $(E = \hbar/T)$. At that instant, collapse occurs, triggering a "moment of awareness" or a conscious event – an event that determines particular configurations of Planck scale experiential geometry and corresponding classical states of nanotubules automata that regulate synaptic and other neural functions. A sequence of such events could provide a forward flow of subjective time and stream of consciousness. Quantum states in nanotubules may link to those in nanotubules in other neurons and glia by tunneling through gap functions,

Table 4.1. The de-coherence relaxation time [1]

| Event | $T$ [ms] | $E$ | $N$ number of aggregates | $T$ [ms] |
|---|---|---|---|---|
| Buddhist moment of awareness nucleons | 13 | $4 \cdot 10^{15}$ | $10^{15}$ | 10 |
| Coherent 40 Hz oscillations | 25 | $2 \cdot 10^{15}$ | $10^{15}$ | 10 |
| EEG alpha rhytm (8 to 12 Hz) | 100 | $10^{14}$ | $10^{14}$ | 1 |
| Libet's sensory threshold | 100 | $10^{14}$ | $10^{14}$ | 1 |

permitting extension of the quantum state through significant volumes of the brain.

Based on $E = \hbar/T$, the size and extension of Orch OR events which correlate with a subjective or neurophysiological description of conscious events can be calculated. In Table 4.1 the calculated $T$ (Penrose-Hameroff) and $\tau$ – equation (4.16) are presented

We shall now develop the generalized quantum heat transport equation for microtubules which also includes the potential term. Thus, we are able to use the analogy of the Schrödinger and quantum heat transport equations. If we consider, for the moment, the parabolic heat transport equation with the second derivative term omitted

$$\frac{\partial T}{\partial t} = \frac{\hbar}{m}\nabla^2 T. \qquad (4.17)$$

If the real time $t \to it/2$, $T \to \Psi$, Eq. (4.17) has the form of a free Schrödinger equation

$$i\hbar\frac{\partial \Psi}{\partial t} = -\frac{\hbar^2}{2m}\nabla^2 \Psi. \qquad (4.18)$$

The complete Schrödinger equation has the form

$$i\hbar\frac{\partial \Psi}{\partial t} = -\frac{\hbar^2}{2m}\nabla^2 \Psi + V\Psi, \qquad (4.19)$$

where $V$ denotes the potential energy. When we go back to real time $t \to 2it$, $\Psi \to T$, the new parabolic heat transport is obtained

$$\frac{\partial T}{\partial t} = \frac{\hbar}{m}\nabla^2 T - \frac{2V}{\hbar}T. \qquad (4.20)$$

Equation (4.20) describes the quantum heat transport for $\Delta t > \tau$. For heat transport initiated by ultra-short laser pulses, when $\Delta t < \tau$ one obtains the generalized quantum hyperbolic heat transport equation

$$\tau\frac{\partial^2 T}{\partial t^2} + \frac{\partial T}{\partial t} = \frac{\hbar}{m}\nabla^2 T - \frac{2V}{\hbar}T. \qquad (4.21)$$

Considering that $\tau = \hbar/mv^2$, Eq. (4.21) can be written as follows:

$$\frac{1}{v^2}\frac{\partial^2 T}{\partial t^2} + \frac{m}{\hbar}\frac{\partial T}{\partial t} + \frac{2Vm}{\hbar^2}T = \nabla^2 T. \qquad (4.22)$$

Equation (4.22) describes the heat flow when apart from the temperature gradient, the potential energy $V$ (is present.)

In the following, we consider one-dimensional heat transfer phenomena, i.e

$$\frac{1}{v^2}\frac{\partial^2 T}{\partial t^2} + \frac{m}{\hbar}\frac{\partial T}{\partial t} + \frac{2Vm}{\hbar^2}T = \frac{\partial^2 T}{\partial x^2}. \qquad (4.23)$$

We seek a solution in the form

$$T(x,t) = e^{1/2\tau}u(x,t). \qquad (4.24)$$

for the quantum heat transport equation (4.23)

After substitution of Eq. (4.24) into Eq. (4.23), one obtains

$$\frac{1}{v^2}\frac{\partial^2 u}{\partial t^2} - \frac{\partial^2 u}{\partial x^2} + qu(x,t) = 0. \qquad (4.25)$$

where

$$q^2 = \frac{2Vm}{\hbar^2} - \left(\frac{mv}{2\hbar}\right)^2 \qquad (4.26)$$

In the following, we consider a constant potential energy $V = V_0$. The general solution of Eq. (4.25) for the Cauchy boundary conditions,

$$u(x,0) = f(x), \qquad \left[\frac{\partial u(x,t)}{\partial t}\right]_{t=0} = F(x), \qquad (4.27)$$

has the form [3]

$$u(x,t) = \frac{f(x-vt) + f(x+vt)}{2} + \frac{1}{2v}\int_{x-vt}^{x+vt}\Phi(x,y,z)dz, \qquad (4.28)$$

where

$$\Phi(x,t,z) = \frac{1}{v}J_0\left(\frac{b}{v}\sqrt{(z-x)^2 - v^2 t^2}\right) + btf(z)\frac{J_0\left(\frac{b}{v}\sqrt{(z-x)^2 - v^2 t^2}\right)}{\sqrt{(z-x)^2 - v^2 t^2}},$$

$$b = \left(\frac{mv^2}{2\hbar}\right) - \frac{2Vm}{\hbar^2}v^2 \qquad (4.29)$$

and $J_0(z)$ denotes the Bessel function of the first kind. Considering equations (4.24), (4.25), (4.26) the solution of Eq. (4.23) describes the propagation of the distorted thermal quantum waves with characteristic lines $x = \pm vt$. We can define the distortionless thermal wave as the wave which preserves the shape in the potential energy $V_0$ field. The condition for conserving the shape can be expressed as

$$q^2 = \frac{2Vm}{\hbar^2} - \left(\frac{mv}{2\hbar}\right)^2 \qquad (4.30)$$

When Eq. (4.30) holds, Eq. (4.31) has the form

$$\frac{\partial^2 u(x,t)}{\partial t^2} = v^2 \frac{\partial^2 u}{\partial x^2}. \qquad (4.31)$$

Equation (4.31) is the quantum wave equation with the solution (for Cauchy boundary conditions (4.27))

$$u(x,t) = \frac{f(x-vt) + f(x+vt)}{2} + \frac{1}{2v}\int_{x-vt}^{x+vt} F(z)dz. \qquad (4.32)$$

It is interesting to observe, that condition (4.30) has an analog in classical theory of the electrical transmission line. In the context of the transmission of an electromagnetic field, the condition $q = 0$ describes the Heaviside distortionless line. Eq. (4.30) – the distortionless condition – can be written as

$$V_0 \tau \approx \hbar, \qquad (4.33)$$

We can conclude, that in the presence of the potential energy $V_0$ one can observe the undisturbed quantum thermal wave in microtubules only when *the Heisenberg uncertainty relation for thermal processes* (4.33) is fulfilled.

The generalized quantum heat transport equation (GQHT) (4.23) leads to generalized Schrödinger equation for microtubules. After the substitution $t \to it/2$, $T \to \Psi$ in Eq. (4.23), one obtains the generalized Schrödinger equation (GSE)

$$i\hbar \frac{\partial \Psi}{\partial t} = -\frac{\hbar^2}{2m}\nabla^2 \Psi + V\Psi - 2\tau\hbar \frac{\partial^2 \Psi}{\partial t^2}. \qquad (4.34)$$

Considering that $\tau = \hbar/mv^2 = \hbar/m\alpha^2 c^2$ ($\alpha = 1/137$) is the fine-structure constant for electromagnetic interactions) Eq. (4.34) can be written as

$$i\hbar \frac{\partial \Psi}{\partial t} = -\frac{\hbar^2}{2m}\nabla^2 \Psi + V\Psi - \frac{2\hbar^2}{m\alpha^2 c^2}\frac{\partial^2 \Psi}{\partial t^2}. \qquad (4.35)$$

One can conclude, that for a time period $\nabla t < \hbar/m\alpha^2 c^2 \approx 10^{-17}$ s the description of quantum phenomena needs some revision. On the other hand, for $\nabla t > 10^{-17}$ in GSE the second derivative term can be omitted and as a result the Schrödinger equation SE is obtained, i.e.

$$i\hbar \frac{\partial \Psi}{\partial t} = -\frac{\hbar^2}{2m}\nabla^2 \Psi + V\Psi \tag{4.36}$$

It is interesting to observe, that GSE was discussed also in the context of the sub-quantal phenomena.

In conclusion a study of the interactions of the attosecond laser pulses with matter can shed light on the applicability of the SE in a study of ultra-short sub-quantal phenomena.

The structure of Eq. (4.25) depends on the sign of the parameter $q$. For quantum heat transport phenomena with electrons as the heat carriers the parameter $q$ is a function of the potential barrier height $V_0$ and velocity $v$.

The initial Cauchy condition

$$u(x,0) = f(x), \qquad \frac{\partial u(x,0)}{\partial t} = g(x), \tag{4.37}$$

and the solution of the Eq. (4.25) has the form [3]

$$u(x,t) = \frac{f(x-vt) + f(x+vt)}{2}$$
$$+ \frac{1}{2v}\int_{x-vt}^{x+vt} g(\varsigma) I_0\left[\sqrt{-q(v^2 t^2 - (x-\varsigma)^2)}\right] d\varsigma \tag{4.38}$$
$$- \frac{(v\sqrt{-q})t}{2}\int_{x-vt}^{x+vt} f(\varsigma)\frac{I_1\left[\sqrt{-q(v^2 t^2 - (x-\varsigma)^2)}\right]}{\sqrt{v^2 t^2 - (x-\varsigma)^2}} d\varsigma.$$

When $q > 0$ Eq. (4.25) is the *Klein – Gordon equation* (K-G), which is well known from applications in elementary particle and nuclear physics.

For the initial Cauchy condition (4.37), the solution of the (K-G) equation can be written as [3]

$$u(x,t) = \frac{f(x-vt) + f(x+vt)}{2}$$
$$+ \frac{1}{2v}\int_{x-vt}^{x+vt} g(\varsigma) J_0\left[\sqrt{q(v^2 t^2 - (x-\varsigma)^2)}\right] d\varsigma \qquad (4.39)$$
$$- \frac{(v\sqrt{q})}{2}\int_{x-vt}^{x+vt} f(\varsigma) \frac{J_0'\left[\sqrt{-q(v^2 t^2 - (x-\varsigma)^2)}\right]}{\sqrt{v^2 t^2 - (x-\varsigma)^2}} d\varsigma.$$

Both solutions (4.38) and (4.39) exhibit the domains of dependence and influence on the *modified Klein-Gordon* and *Klein-Gordon equation.* These domains, which characterize the maximum speed at which a thermal disturbance can travel are determined by the principal terms of the given equation (i.e., the second derivative terms) and do not depend on the lower order terms. It can be concluded that these equations and the wave equation (for *m* = 0) have identical domains of dependence and influence.

4.3 Heat transport with Casimir potential included

Vacuum energy is a consequence of the quantum nature of the electromagnetic field, which is composed of photons. A photon of frequency $\omega$ has an energy $\hbar\omega$, where $\hbar$ is Planck constant. The quantum vacuum can be interpreted as the lowest energy state (or ground state) of the electromagnetic (EM) field which result when all charges and currents have been removed and the temperature has been reduced to absolute zero. In this state no ordinary photons are present. Nevertheless, because the electromagnetic field is a quantum system the energy of the ground state of the EM is not zero. Although the average value of the electric field $\langle E \rangle$ vanishes in ground state, the Root Mean Square of the field $\langle E^2 \rangle$ is not zero. Similarly the $\langle B^2 \rangle$ is not zero. Therefore the electromagnetic field energy $\langle E^2 \rangle + \langle B^2 \rangle$ is not equal zero. A detailed theoretical calculation tells that EM energy in each mode of oscillation with frequency $\omega$ is 0.5 $\hbar\omega$, which equals one half of amount energy that would be present if a single "real" photon of that mode were present. Adding up 0.5 $\hbar\omega$ for each possible states of the electromagnetic field result in a very large number for the vacuum energy $E_0$ in a quantum vacuum

$$E_0 = \sum_i \frac{1}{2}\hbar\omega_i. \tag{4.40}$$

The resulting vacuum energy $E_0$ is *infinity* unless a high frequency cut off is applied.

Inserting surfaces into the vacuum causes the states of the EM field to change. This change in the states takes place because the EM field must meet the appropriate boundary conditions at each surface. The surfaces alter the modes of oscillation and therefore alter the energy density of the lowest state of the EM field. In actual practice the change in $E_0$ is

$$\Delta E_0 = E_0 - E_S \tag{4.41}$$

where $E_0$ is the energy in empty space and $E_S$ is the energy in space with surfaces, i.e.

$$\Delta E_0 = \frac{1}{2}\sum_n^{\substack{\text{empty}\\\text{space}}} \hbar\omega_n - \frac{1}{2}\sum_i^{\substack{\text{surface}\\\text{present}}} \hbar\omega_i. \tag{4.42}$$

As an example let us consider a hollow conducting rectangular cavity with sides $a_1$, $a_2$, $a_3$. In this case for uncharged parallel plates with an area $A$ the attractive force between the plates is, $F_{att} = -\frac{\pi^2 \hbar c}{240 d^4} A,$ (4.43) where $d$ is the distance between plates. The force $F_{att}$ is called the parallel plate Casimir force, which was measured in three different experiments .

Recent calculations show that for conductive rectangular cavities the vacuum forces on a given face can be repulsive (positive), attractive (negative) or zero depending on the ratio of the sides .

The first measurement of repulsive Casimir force was performed by Maclay . For a distance of separation of $d \sim 0.1$ μm the repulsive force is of the order of 0.5 μN (micronewton)– for cavity geometry. In March 2001, scientist at Lucent Technology used the attractive parallel plate Casimir force to actuate a MEMS torsion device [4.6]. Other MEMS (MicroElectroMechanical System) have been also proposed

Standard Klein – Gordon equation is expressed as:

$$\frac{1}{c^2}\frac{\partial^2 \Psi}{\partial t^2} - \frac{\partial^2 \Psi}{\partial x^2} + \frac{m^2 c^2}{\hbar^2}\Psi = 0. \tag{4.44}$$

In equation (4.44) $\Psi$ is the relativistic wave function for particle with mass $m$, $c$ is the speed of light and $\hbar$ is Planck constant. In case of massless particles $m = 0$ and Eq. (4.44) is the Maxwell equation for photons. As was shown by Pauli and Weiskopf, the Klein – Gordon equation describes spin, – 0 bosons, because relativistic quantum mechanical equation had to allow for creation and annihilation of particles.

In the monograph by J. Marciak – Kozłowska and M. Kozłowski [1] the generalized Klein – Gordon thermal equation was developed

$$\frac{1}{v^2}\frac{\partial^2 T}{\partial t^2} - \nabla^2 T + \frac{m}{\hbar}\frac{\partial T}{\partial t} + \frac{2Vm}{\hbar^2} = 0. \tag{4.45}$$

In Eq. (4.45) $T$ denotes temperature of the medium and $v$ is the velocity of the temperature signal in the medium. When we extract the highly oscillating part of the temperature field,

$$T = e^{-\frac{t\omega}{2}} u(x,t), \tag{4.46}$$

where $\omega = \tau^{-1}$, and $\tau$ is the relaxation time, we obtain from Eq. (4.42) (1D case)

$$\frac{1}{v^2}\frac{\partial^2 u}{\partial t^2} - \frac{\partial^2 u}{\partial x^2} + qu(x,t) = 0, \tag{4.47}$$

where

$$q = \frac{2Vm}{\hbar^2} - \left(\frac{mv}{2\hbar}\right)^2. \tag{4.48}$$

When $q > 0$ equation (4.43) is of the form of the Klein – Gordon equation in the potential field $V(x, t)$. For $q < 0$ Eq. (4.47) is the modified Klein – Gordon equation.

Considering the existence of the attosecond laser with $\Delta t = 1$ as $= 10^{-18}$s, Eq. (4.47) describes the heat signaling for thermal energy transport induced by ultra-short laser pulses. In the subsequent we will consider the heat transport when $V$ is the Casimir potential. As was shown in paper [4.8] the Casimir force, formula (4.43), can be repulsive sign $(V) = +1$ and attractive sign $(V) = -1$. For attractive Casimir force, $V < 0$, $q < 0$ (formula (4.44)) and equation (4.43) is the modified K-G equation. For repulsive Casimir force $V > 0$ and $q$ can be positive or negative.

As was shown by J. Maclay for different shapes of cavities, the vacuum Casimir force can change sign. Below we consider the propagation of a thermal wave within parallel plates. For Cauchy initial condition:

$$u(x,0) = 0, \qquad \frac{\partial u(x,0)}{\partial t} = g(x)$$

the solution of Eq. (4.47) has the form

$$u(x,t) = \frac{1}{2v} \int_{x-vt}^{x+vt} g(\zeta) J_0\left(\sqrt{q(v^2 t^2 - (x-\zeta)^2)}\right) d\zeta \qquad \text{for} \qquad q > 0 \qquad (4.50)$$

and

$$u(x,t) = \frac{1}{2v} \int_{x-vt}^{x+vt} g(\zeta) I_0\left(\sqrt{-q(v^2 t^2 - (x-\zeta)^2)}\right) d\zeta \qquad \text{for} \qquad q < 0 \qquad (4.51)$$

References

[1] M Kozłowski, J Marciak-Kozłowska, Thermal processes using attosecond laser pulses, Springer USA, 2006

[2] D Jou et al., Extended irreversible thermodynamics, Springer, 2001

[3] E Zauderer, Partial differential equations of applied mathematics, Wiley 1989

61